%% file: 0Paperid1633.tex
\documentclass[conference]{IEEEtran}
\IEEEoverridecommandlockouts

\usepackage{cite}
\usepackage{amsmath,amssymb,amsfonts}
\usepackage{algorithmic}
\usepackage{graphicx}
\usepackage{xcolor}
\usepackage[ruled,linesnumbered]{algorithm2e}
\usepackage{multirow}
\usepackage{color, xspace}
\usepackage{enumitem}
\usepackage{CJKutf8}
\usepackage{url}

\def\BibTeX{{\rm B\kern-.05em{\sc i\kern-.025em b}\kern-.08em
    T\kern-.1667em\lower.7ex\hbox{E}\kern-.125emX}}
\begin{document}
\begin{CJK}{UTF8}{gbsn}
\title{Agent4Ranking: Semantic Robust Ranking via Personalized Query Rewriting Using Multi-agent LLM
}
\author{
    \IEEEauthorblockN{Xiaopeng Li$^{\dagger}$, Lixin Su$^{\ddagger}$, Pengyue Jia$^{\dagger}$, Xiangyu Zhao$^{\dagger}$, Suqi Cheng$^{\ddagger}$, Junfeng Wang$^{\ddagger}$, Dawei Yin$^{\ddagger}$}
    \IEEEauthorblockA{$^\dagger$ City Univerisity of HongKong}
    \IEEEauthorblockA{$^\ddagger$ Baidu Inc.}
    \IEEEauthorblockA{\{xiaopli2-c, jia.pengyue\}@my.cityu.edu.hk, xianzhao@cityu.edu.hk,\\ \{sulixin, chengsuqi, wangjunfeng\}@baidu.com, yindawei@acm.org}
}

\maketitle

\begin{abstract}
Search engines are crucial as they provide an efficient and easy way to access vast amounts of information on the internet for diverse information needs.
User queries, even with a specific need, can differ significantly. Prior research has explored the resilience of ranking models against typical query variations like paraphrasing, misspellings, and order changes. Yet, these works overlook how diverse demographics uniquely formulate identical queries.
For instance, older individuals tend to construct queries more naturally and in varied order compared to other groups. This demographic diversity necessitates enhancing the adaptability of ranking models to diverse query formulations.
To this end, in this paper, we propose a framework that integrates a novel rewriting pipeline that rewrites queries from various demographic perspectives and a novel framework to enhance ranking robustness. To be specific, we use Chain of Thought (CoT) technology to utilize Large Language Models (LLMs) as agents to emulate various demographic profiles, then use them for efficient query rewriting, and we innovate a robust Multi-gate Mixture of Experts (MMoE) architecture coupled with a hybrid loss function, collectively strengthening the ranking models' robustness. Our extensive experimentation on both public and industrial datasets assesses the efficacy of our query rewriting approach and the enhanced accuracy and robustness of the ranking model. The findings highlight the sophistication and effectiveness of our proposed model.
\end{abstract}

\input{1Introduction}

\input{2Preliminary}

\input{3Methodology}
\input{4Experiment}

\input{5Relatedwork}
\input{6Conclusion}
\bibliographystyle{IEEEtran}
\bibliography{mybibfile}

\vspace{12pt}

\end{CJK}
\end{document}

%% file: 1Introduction.tex
\section{Introduction}

Search engines, pivotal in Computer Science and Data Management, focus on efficiently searching and retrieving relevant information from extensive data repositories~\cite{ntoulas2004s}. A central challenge in search engine technology is addressing the ranking problem, which involves prioritizing search results according to their relevance to user queries. Ranking models have evolved significantly, encompassing three main types: traditional probabilistic models~\cite{salton1975vector,robertson1976relevance}, which falter in large-scale applications due to their keyword-centric design; neural rankers~\cite{dai2019deeper, dai2018convolutional, guo2016deep} that leverage deep learning for enhanced performance; and recent pre-trained models~\cite{gu2020speaker,ma2021b} capable of understanding complex queries with limited data. Most of this research has principally focused on evaluating and enhancing the effectiveness of ranking models. However, the robustness of these ranking models, which refers to the stability of document rankings under the influence of perturbed queries, has received relatively less attention.

In practical search scenarios, the robustness of search systems in ranking is of paramount importance. Ideally, a robust search engine should return consistent results for semantically consistent queries~\cite{dash2003consistency}. However, a lack of robustness can lead to significant variability in results, adversely affecting the user experience. Existing studies have explored the impact of query variations on robustness of the search engine pipeline. For example, Penha et al.\cite{penha2022evaluating} examine the effects of variations like misspellings and paraphrasing, identifying a notable 20\% decline in retrieval effectiveness, indicating current systems' robustness deficiencies. CAPOT~\cite{campos2023noise} employs contrastive learning to address noisy query impacts including typos and misspellings. Zhuang et al. investigate the use of Character-Bert and self-teaching techniques~\cite{zhuang2022characterbert} for effectively handling typo-laden queries. Nevertheless, the majority of these studies predominantly concentrate on the retrieval process lacking exploration of the ranking procedure, more importantly, these studies heuristically define types of query variations and tackle insufficient robustness. Practically, the search engine often confronts mixed query variations composed of multiple types of query alterations. 
Intuitively, different demographic groups have the same information requirement. But limited by their different personal knowledge background, they express different queries. Therefore, a more direct way is to simulate different demographic groups of people to rewrite queries based on the information needs.

\begin{figure}[ht]
  \centering
 \includegraphics[width = 0.8\linewidth]{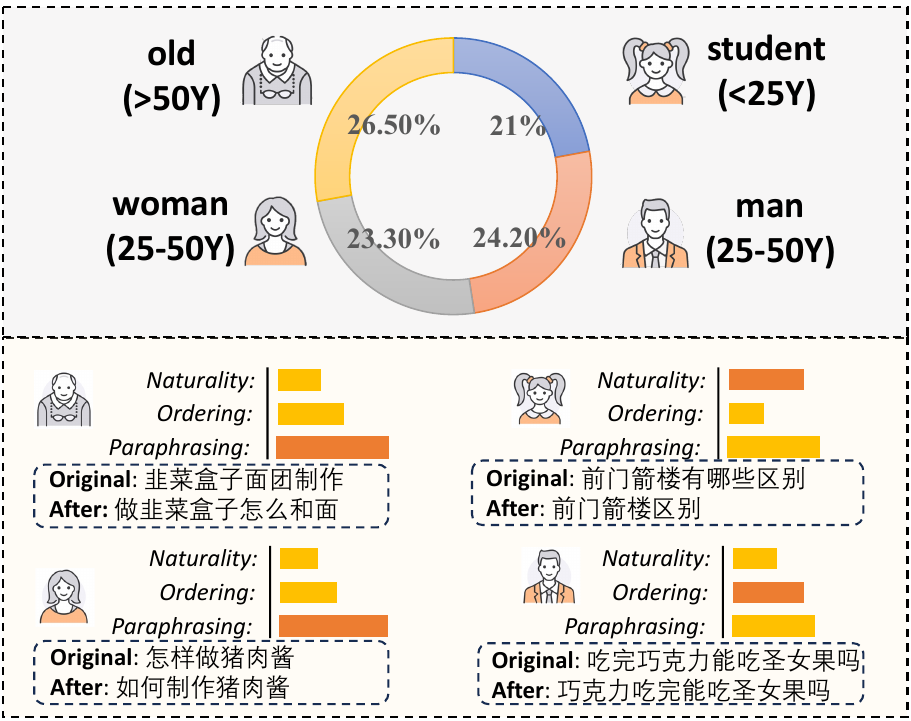}
 \caption{Up: Four demographic group statistics by CNNIC~\cite{statistic2021cnnic}. Down: Taxonomy of query variations statistics for four groups. All the queries are from the industrial dataset in Baidu. The Red Bar indicates the most significant feature of this group.}
 \label{fig:statistics}
\end{figure}

To tackle the aforementioned issues, this paper proposes to investigate the application of Large Language Models (LLMs) in query rewriting from demographic perspectives, aiming to fortify the robustness of ranking models for different groups of users. LLMs have emerged as formidable tools, noted for their exceptional contextual awareness and generative abilities, are particularly effective in query rewriting, ensuring semantic integrity and simultaneously boosting model robustness~\cite{jia2023mill}. However, it is not easy to apply LLMs into query rewriting in demographic perspectives. 
Firstly, how to accurately rewrite queries from various perspectives using LLMs such as ChatGPT. These models are notable for their contextual understanding and semantic comprehension. Nevertheless, the unpredictability in their outputs often leads to generation inaccuracies or "hallucinations"~\cite{hao2022cgf}. The complexity of crafting prompts that guide LLMs to produce contextually relevant and diverse query rewrites exacerbates this issue.
Secondly, how to enhance the robustness of ranking models via multiple query variants. Current literature primarily addresses single perspective variants, such as misspellings or ordering, and often neglects the complexity of mixed query rewrites in practical scenarios~\cite{penha2022evaluating}. Consequently, the development of a robust model capable of handling a variety of semantically similar queries, and consistently producing similar ranking outcomes without sacrificing accuracy is of paramount importance.

Recent advancements in query rewriting using LLMs predominantly fall into two categories.
The first is the corpora-incorporated approach, exemplified by HyDE~\cite{gao2022precise}, which generates hypothetical documents from original queries for enhanced retrieval, and Query2doc~\cite{wang2023query2doc}, which employs few-shot learning in LLMs to create pseudo documents for query rewriting. 
Shen et al. proposed a  method~\cite{shen2023large} that integrates potential relevant documents into LLM prompts to modify queries.
The second category, generative rewriting, leverages LLMs' language understanding and pre-existing knowledge for query regeneration, such as using prompting~\cite{jia2023mill,alaofi2023can,jagerman2023query} and domain-specific fine-tuning methods~\cite{yu2022generate,izacard2020leveraging} for the rewriting task. 
Nevertheless, these approaches generally focus on single-perspective rewriting and lack an extensive exploration of multi-perspective rewriting. They also fall short in addressing the quality of rewritten queries, particularly in terms of hallucination effects and the strategies for their mitigation. Moreover, there's a gap in exploring how to systematically utilize these rewritten queries to jointly enhance the robustness of ranking models.

Addressing the aforementioned challenges, in this paper, we introduce an innovative pipeline for query rewriting and a framework for enhancing robustness for ranking within search engines. To improve the effectiveness of query rewriting from varied semantic perspectives, we harness the potential of LLMs to simulate four distinct agent roles—woman, man, student, and elder—and rephrase queries accordingly. We carefully chose these four roles according to the statistics from the government~\cite{statistic2021cnnic}, which cover the majority of Chinese Internet users, shown in Figure~\ref{fig:statistics}. We integrate Chain of Thought (CoT) technology and implement rigorous query verification procedures to counteract hallucination effects and enhance the precision of query generation. This approach is encapsulated within a cyclic generation framework that evaluates and iteratively refines the query until it meets predefined quality standards. According to our results analysis, our methodology demonstrates exceptional performance and closely aligns with the character's personality. Moreover, to reinforce the robustness of the ranking model, we introduce a Robust MMoE structure. This structure dynamically identifies semantic commonalities across various rewritten queries, facilitating a more stable ranking process. Additionally, we develop a novel loss function that promotes robustness by leveraging Jensen-Shannon divergence to measure distribution variations across different agent perspectives, thus enhancing the robustness of the result. The paper's primary contributions are as follows:

\begin{itemize} [leftmargin=*]
\item  We introduce a novel query rewriting approach utilizing Large Language Models (LLMs) in various agent roles. This method is supplemented by a query validation procedure that rigorously assesses and iteratively refines the queries, enhancing precision until they conform to our high standards;

\item In order to enhance the robustness of the ranking model, we design a robust ranking model that incorporates a Mixture-of-Experts (MoE) structure with multiple adapters. This architecture is adept at capturing semantic similarities, thereby fortifying the model's robustness. Concurrently, we also develop a novel loss function that strategically enhances model robustness by constraining output distributions.

\item We execute our experiments on two distinct datasets: a publicly available dataset and an industrial dataset from Baidu. Our extensive experimental setup encompasses query evaluation, effectiveness assessments, and robustness performance tests, which collectively affirm the strengths of our proposed framework.

\end{itemize}

%% file: 2Preliminary.tex
\begin{figure*}[htbp]
  \centering
 \includegraphics[width = 0.8\linewidth]{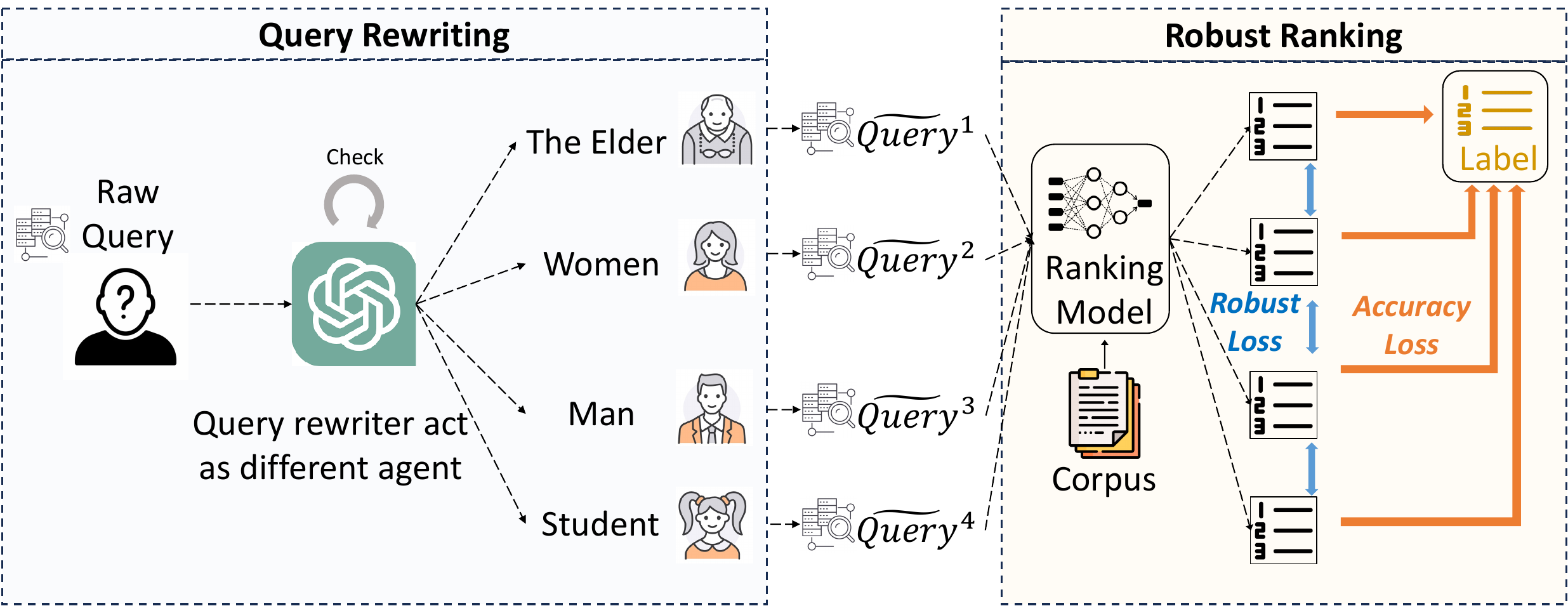}
 \caption{An illustration of the overall architecture of our proposed model. 
 }
 \label{fig:Framework}
 \vspace*{-3mm}
\end{figure*}
\section{Preliminary}
In this section, we delineate the principal notations and the mathematical formulation of our research problem, which encompasses query rewriting and ranking tasks.

Consider a set of original user queries, $\mathcal{Q} = {q_0,...,q_i,...,q_n}$, with $n$ representing the total number of queries. Our objective is to rewrite these queries from $K$ distinct perspectives. We define the $k$-th rewriting function as $\mathcal{F}{\theta{k}}$, where $\theta_k$ parameterizes the rewriting process. The set of rewritten queries corresponding to the $k$-th perspective is represented as $\widetilde{\mathcal{Q}}^{k} = {\widetilde{q}^{k}_0,...,\widetilde{q}^{k}_i,...,\widetilde{q}^{k}n}$. This process can be formalized as
\begin{equation*}
\mathcal{F}{\theta_k}: \mathcal{Q} \rightarrow \widetilde{\mathcal{Q}}^{k},~k\in[1,...,K]
\end{equation*}

Suppose each query $q_i$ in $\mathcal{Q}$ is associated with a document list $\mathcal{D}_i = \{ d_{i,1},d_{i,2},...,d_{i,n_{q_i}} \}$, and for the corresponding list of labels is $\boldsymbol{Y}_i = \{ \boldsymbol{y}_{i,1},\boldsymbol{y}_{i,2},...,\boldsymbol{y}_{i,n_{q_i}} \}$, where $n_{q_i}$ represents the length of list $\mathcal{D}_i$ and $d_{i,j}$ denotes the $j$-th document in document list $\mathcal{D}_i$ for query $q_i$. We consider a ranking model as $\mathcal{R}$, which is trained on the samples $\{q_i,\mathcal{D}_i,\boldsymbol{Y}_i\}_{i=1}^t$ drawn from the training dataset distribution $\mathcal{G}$. The ranking process is defined as producing a permutation $\pi(q_i, \mathcal{D}_i, \mathcal{R})$ for the given query $q_i$ and the corresponding document list $\mathcal{D}_i$ using ranking model $\mathcal{R}$.

The evaluation of the ranking model's effectiveness can be expressed as follows
\begin{equation}
\mathbb{E}_{(q_c,\mathcal{D}_c,\boldsymbol{Y}_c )} \sim \mathcal{M} \left( \pi\left( q_c,\mathcal{D}_c,\mathcal{R} \right), \boldsymbol{Y}_c \right)
\end{equation}
Here, $\mathcal{M}$ represents the metric for evaluating effectiveness, and the tuple $(q_c,\mathcal{D}_c,\boldsymbol{Y}_c )$ signifies the test queries, their associated document lists, and ranking labels. Notably, the test dataset is derived from the same distribution $\mathcal{G}$ as the training set. The ranking model's performance is quantified by the mean effectiveness across test datasets.

The robustness of the ranking model is quantified as
\begin{equation}
\begin{split}
\mathbb{V}_{(q_c,\mathcal{D}_c)} \sim \mathcal{M} ([\pi^k(q^k_c,\mathcal{D}_c,\mathcal{R})]),k \in [1,..,K]
\end{split}
\end{equation}
In this case, $\pi^k (q^k_c,\mathcal{D}_c,\mathcal{R})$ indicates the ranking label list for the $k$-th rewritten query and its corresponding documents using the ranking model. The robustness is assessed by the variance in ranking permutations for different rewritten queries.

%% file: 3Methodology.tex
\section{Methodology}
In this section, we will offer a comprehensive exposition of the methodology employed in our study. At first, we will present a conceptual overview of the proposed architectural framework. This is followed by an in-depth examination of two constituent subtasks, query rewriting and robust ranking, which are explicated sequentially.

\begin{figure*}[htbp]
  \centering
 \includegraphics[width = 0.8\linewidth]{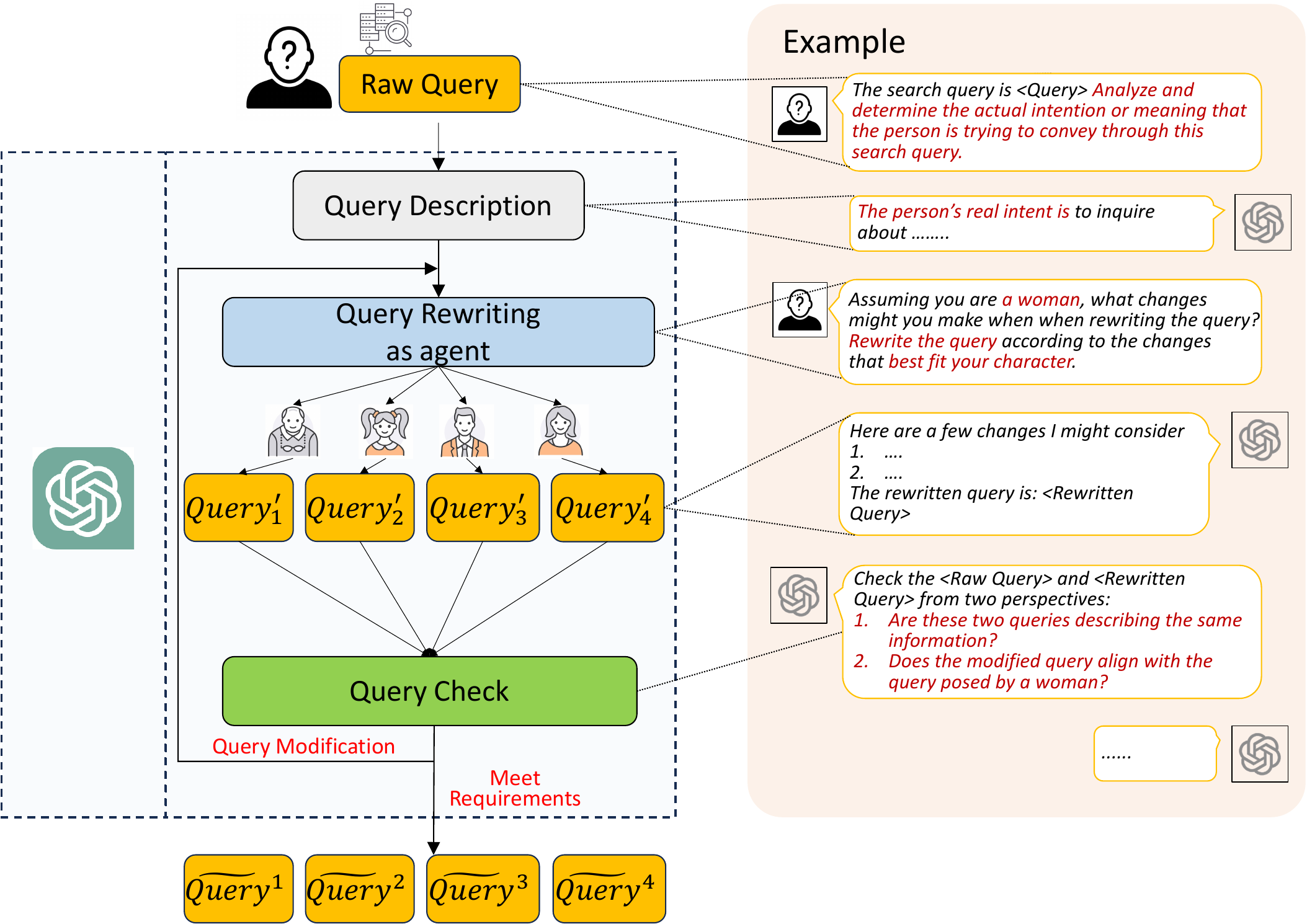}
 \caption{An illustration of the rewriting pipeline of our proposed model. 
 }
 \label{fig:Rewriting}
 \vspace*{-3mm}
\end{figure*}

\subsection{Framework Overview}
The framework of our study is shown in Figure~\ref{fig:Framework}. Our study encompasses two distinct yet interconnected subtasks. The first task is query rewriting. Unlike traditional methods that solely emphasize the quality of rewritten queries, we extend our consideration to rewriting from various semantic viewpoints. The recent developments in the field of LLM agents have provided an opportunity that allows different agents acting in diverse roles to do different tasks. Our research explores how these different roles agents can rewrite queries in a manner that aligns with corresponding demographic group characteristics. 
The second subtask is robust ranking. The current ranking model does not return consistent results when faced with queries with the same semantics, which proves the non-robustness of the existing ranking model. This is primarily due to the ranking module's weak semantic comprehension between queries and documents. To enhance robustness, we raise a robust MMoE ranking model to effectively capture shared semantic information across different queries. Additionally, we devise a loss function, imposing constraints on accuracy and robustness at the same time, increasing robustness while ensuring ranking accuracy. 
\begin{itemize}[leftmargin=*]
\item \textbf{Query Rewriting}: 
The optimization of query rewriting commences with the application of ChatGPT-3.5, where prompts are carefully structured to distill the core intent of user queries, thereby revealing their underlying information needs. Subsequently, distinct personas representing four demographic groups — the elderly, middle-aged women, middle-aged men, and students — are assigned to the LLM. We carefully choose these four roles to pretend to be real users on the Internet. The LLM is then tasked with performing the query rewriting under these specified roles. Nonetheless, due to the hallucination limitations inherent in LLMs, the resultant queries may substantially diverge from the original semantics. To counteract this issue, we implement a critical evaluation of the rewrites, including two aspects: Semantic fidelity to the original query and consistency of the rewritten tone with the designated persona. Queries not meeting our standards are subjected to additional refinement, employing an iterative approach until the desired quality threshold is attained. This cycle of evaluation and refinement is sustained until the rephrased queries align semantically with the original intent and correspond with the assigned persona's tone, thereby ensuring the final output's utmost quality and appropriateness.

\item \textbf{Robust Ranking}: 
To augment the robustness of document ranking amidst semantically similar queries, our approach encompasses two primary dimensions. First, we aim to bolster robustness structurally. The prevalent issue in existing cross-encoders is their limited capacity to discern similar semantic content across varied queries. To overcome this, we introduce the Robust MMoE, an innovative structure integrated atop the transformer model. This model consists of two main components: a query-specific expert and a query-shared expert, designed to capture both query-independent and shared semantic representations, respectively. Second, concerning the loss function, we advocate a loss function designed to optimize both accuracy and robustness. The robust loss segment computes the Jensen–Shannon divergence between the document ranking scores of different queries, thereby enhancing robustness. Simultaneously, the accuracy loss segment focuses on maintaining the precision of the ranking, thus ensuring accuracy.

\end{itemize}

\subsection{Query Rewriting}

This subsection delves into the detailed methodology of our rewriting pipeline. 
Large Language Models (LLMs) have recently become prominent as effective query rewriters in search engines\cite{ma2023query,anand2023query}. While most existing research concentrates on single perspective variants, such as misspellings or typos~\cite{zhuang2022characterbert}, the complexity of multi-semantic query rewrites in real-world scenarios is often overlooked~\cite{penha2022evaluating}. In our study, we utilize LLMs as agents adopting diverse roles to skillfully rewrite queries from multiple semantic perspectives, and we enhance the quality of these rewrites through an iterative process. The workflow of this process is illustrated in Figure~\ref{fig:Rewriting} for a comprehensive understanding.

The essence of the query rewriting task lies in comprehending the user's intent and discerning the actual information they seek, which is crucial for effective query rewriting. To achieve this, we leverage the advanced semantic understanding capabilities of LLMs to accurately ascertain the underlying information needs within the queries. In practical terms, given an initial set of raw queries denoted as $\mathcal{Q}^r$, we utilize prompt engineering techniques to craft prompt~(a), as detailed in Table~\ref{tab:rewriting_propmt}. This approach is instrumental in extracting pertinent details from the queries and precisely capturing the user's authentic intent.

Following the initial extraction of information, we assign the LLM to diverse agent roles for query rewriting. To facilitate this, prompt~(b) is provided to guide the LLMs. These agent roles are meticulously crafted to mirror four distinct demographic groups: middle-aged men, middle-aged women, the elderly, and students, with the resultant queries represented as $\mathcal{Q}'_k, k\in[1,2,3,4]$. According to the demographic investigation on the Internet~\cite{statistic2021cnnic}, these four groups were selected to offer a balanced representation across age spectrums, categorizing users over 50 years as old and those under 25 as student, with the remaining demographics segmented by gender, results are shown in Figure~\ref{fig:statistics}. Additionally, these groups showcase unique semantic expressions in their queries, encompassing aspects of naturality, ordering, and paraphrasing~\cite{alaofi2023can}. For example, the old prefer colloquial language, whereas students, more adept with search tools, tend to use keyword-driven queries. Thus, each demographic group exhibits distinctive query expression patterns influenced by their age and gender characteristics.

Upon obtaining the queries reformulated by four distinct agent roles, conducting a rigorous quality assessment is imperative. This evaluation step is crucial to ascertain the fidelity of the regenerated queries. Given that the issue of hallucinations in LLMs is not entirely resolved, there remains a possibility of some queries failing to align with the set standards. For example, some queries might undergo substantial semantic alteration, deviating from their original intent, or may not be appropriately rephrased in accordance with the assigned agent role.

To mitigate these concerns, we evaluate the rewritten queries from two different perspectives. Firstly, it assesses the semantic similarity between the original and rewritten queries. Secondly, it evaluates the adherence of the rewritten queries to their designated agent roles. For both evaluations, a three-tiered scoring mechanism is implemented: fully meeting the requirements, partially meeting the requirements, and not meeting the requirements. Details of the corresponding prompts are presented as prompt~(c) in Table~\ref{tab:rewriting_propmt}.

When queries fall short of the set criteria, we will redirect them to the query rewriting phase with modified instructions for the LLMs. In cases where the queries are of substandard quality, particularly regarding semantic fidelity, we mandate a closer alignment of the regenerated queries with the original semantics, as elaborated in prompt~(d). In instances where the query inadequately conforms to the designated agent role, we craft the prompt to underscore the need for enhanced adherence to the agent role, as delineated in prompt~(e). To effectively tackle both quality and role conformity challenges simultaneously, a tailored prompt~(f) is formulated for comprehensive mitigation.

\begin{table}[htbp]
\caption{Prompts of query rewriting for public dataset}
\begin{center}
\begin{tabular}{ p{1.7cm} | p{5.5cm} }
\hline
\textbf{Methods} & \textbf{Prompts} \\
\hline
\multirow{2}{1.5cm}[-0.6cm]{\centering Queries  Generation} & (a). The search query is \{query\}. Please analyze and determine the actual intention or meaning that the person is trying to convey through this search query.\\ \cline{2-2}
 & (b). Assuming you are a woman, what changes might you make when rewriting the query? Please rewrite the query to align it with your role.\\
\hline
\multirow{1}{1.5cm}[-1.2cm]{\centering Queries Check} & (c). The original query is: \{query\}. The rephrased query is: \{rewriting query\}. Evaluate the following:1. Are these two queries describing the same information? 2. Does the modified query align with the query posed by \{agent\}? Assign judgment scores of -1, 0, or 1. A score of -1 implies no match, 0 suggests an approximate match, and 1 indicates an exact match.\\
\hline
\multirow{1}{1.5cm}[-1.7cm]{\centering Queries Modification} & (d). Assuming you are an \{agent\}, please rephrase the query in accordance with your role while preserving the original meaning of the question. \\
\cline{2-2}
~ & (e). Assuming you are an \{agent\}, please rephrase the query according to your role and rewrite it more in line with the character's attributes. \\
\cline{2-2}
~ & (f). Assuming you are an \{agent\}, please rephrase the question consistent with your role, maintaining the essence of the original query and aligning it with the character's attributes.\\
\hline
\end{tabular}
\label{tab:rewriting_propmt}
\end{center}
\end{table}

This iterative process continues until the generation of queries that satisfy the established requirements, denoted as $\widetilde{\mathcal{Q}}^k$. The algorithm for the entire rewriting process is depicted in Algorithm~\ref{alg:Rewriting}.

This process is repeated until the queries that meet the requirements are generated, denoted as $\widetilde{\mathcal{Q}}^k$ The algorithm for the entire rewriting process is depicted in Algorithm~\ref{alg:Rewriting}.

\begin{algorithm}[t]
\begin{footnotesize}
\caption{\label{alg:Rewriting} The Algorithm for rewriting.}\
\KwIn{Input queries $\mathcal{Q}^r$, prompts $\mathcal{P}_a,\mathcal{P}_b,\mathcal{P}_c,\mathcal{P}_d,\mathcal{P}_e$}, agent list $K$.

\KwOut{Rewritten queries $\widetilde{\mathcal{Q}}^k, k \in K$}
 \ForEach{$q_i \in \mathcal{Q}^r$}{
 Information extraction $e_i = \mathcal{P}_a(q_i)$\;
 \ForEach{ \text{agent} $k \in K$}{
 Persona rewriting $q'_{i_k} = \mathcal{P}_b(e_i, k)$\;
 Get quality and persona score $s_0,s_1$ via query check $s_0,s_1 = \mathcal{P}_c(q'_{i_k}, q_{i}, k)$\;

\While{$s_0<0$ or $s_1<0$}{
\If{$s_0<0$ and $s_1>=0$}{
Query regeneration $q'_{i_k} = \mathcal{P}_d(e_i, q_{i}, q'_{i_k}, k)$\;
   }
\eIf{$s_1<0$ and $s_0>=0$}{
Query regeneration $q'_{i_k} = \mathcal{P}_e(e_i, q_{i}, q'_{i_k}, k)$\;
}
{
Query regeneration $q'_{i_k} = \mathcal{P}_f(e_i, q_{i}, q'_{i_k}, k)$\;
}
Query Check $s_0,s_1 = \mathcal{P}_c(q'_{i_k}, q_{i}, k)$\;
}
Add $q'_{i_k}$ to list $\widetilde{q}^k$\;
}
Add $q^k$ to $\widetilde{\mathcal{Q}}^k$\;
}
\end{footnotesize}
\end{algorithm}  
\vspace*{-3mm}

\subsection{Robust Ranking}
In this subsection, we will introduce how to enhance the robustness of the existing ranking model.
\begin{figure*}[htbp]
  \centering
 \includegraphics[width = 0.8\linewidth]{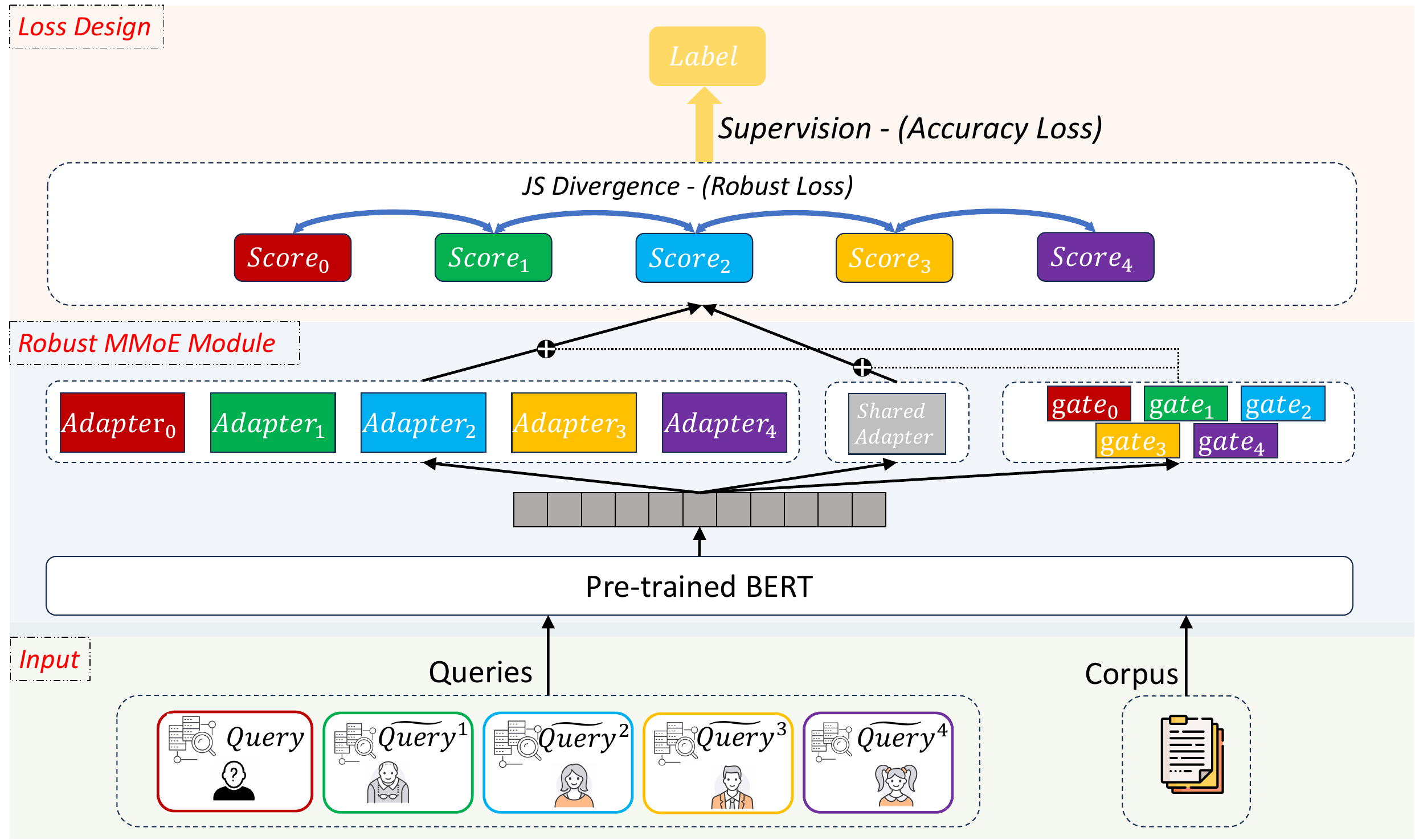}
 \caption{An illustration of the robust ranking architecture of our proposed model. 
 }
 \label{fig:Robust}
 \vspace*{-3mm}
\end{figure*}
The conventional cross-encoder~\cite{reimers2019sentence} takes in a pair of texts, typically a search query and a document, and encodes text into a single contextual vector. This vector is then compressed into a singular value, serving as a metric to assess the relationship or similarity between the query and the document. Despite its widespread usage, we argue that the existing cross-encoder framework exhibits insufficient robustness for two primary reasons. One major drawback lies in the insufficient capacity to extract shared semantic information effectively. In instances where inputs are semantically identical, the current cross-encoder architecture fails to capture shared semantic nuances adequately, thus resulting in unrobust ranking outcomes and hindering the model's overall performance. Another critical factor contributing to the inadequate robustness is the lack of stringent constraints in the training process. As a consequence, the model fails to generalize effectively across semantically similar instances, leading to a lack of robustness in real-world ranking scenarios.

In response to the above two points, our approach begins with enhancements in structure and loss function design to boost robustness. Specifically, we have developed a robust Multi-gate Mixture of Experts (MMoE)-adapter structure coupled with a robust loss function, collaboratively designed to augment the ranker's robustness.

The specific structure is shown in the Figure~\ref{fig:Robust}. In the following content, we will first introduce the input, followed by detailed discussions on the two primary modules: the robust Mixed Model of Experts (MMoE)-adapter and the module's loss design.

\subsubsection{Data Input}
For different agent queries $\widetilde{\mathcal{Q}}^{k}, k\in[1,...,4]$, where $k$ denotes $k$-th agent, and the raw query list $Q^r$. We have a document list $\mathcal{D}_i$ for $i$-th query that needed to be ranked. We combine the above data into candidate pairs and formulate the input pairs as follows
\begin{equation*}
\text{Input}:~<q^r_i, d_{i,j}>, <q^1_i, d_{i,j}>,...,<q^4_i, d_{i,j}>
\end{equation*}
where $q^r_i \in Q^r$, $ q^k_i \in \widetilde{\mathcal{Q}}^{k}$ and $d_{i,j}$ represents $j$-th document in document list $\mathcal{D}_i$.
For each pair, we model them as the format $\boldsymbol{s}_{i,j} = <[\text{CLS}]~q_i~[\text{SEP}]~d_{i,j}~[\text{SEP}]>$ and feed them into the pre-trained transformer model to derive the intermediate dense representation for each pair, formulate as follows
\begin{equation}
\label{equ:PLM}
\begin{aligned}
\boldsymbol{e}^r_i &= \text{PLM}(\boldsymbol{s}^r_{i,j})\\
\boldsymbol{e}^k_i = \text{PLM}&(\boldsymbol{s}^k_{i,j}), k\in [1,2,3,4]
\end{aligned}
\end{equation}
where $\boldsymbol{e}^r_i$ denotes the raw query-doc representation and $\boldsymbol{e}^k_i$ denotes $k$-th agent rewritten query-doc representation. We selected three different transformer models for the experiment: BERT~\cite{devlin2018bert}, RoBERTa~\cite{liu2019roberta}, and ERNIE~\cite{sun2020ernie}. To achieve optimal performance, preliminary training on the training dataset is conducted for these PLMs, followed by fine-tuning for the whole model to enhance the robustness. Detailed descriptions of the whole model design and experimental procedures will be presented in the subsequent section.

\subsubsection{Robust MMoE Module}

In this part, we present the methodology for designing an MMoE-adapter structure as an extension of an existing model to enhance robustness under multiple consistent semantic query inputs.

The Multi-gate Mixture-of-Experts (MMoE) model, originally introduced by Ma et al.~\cite{ma2018modeling}, tackles the challenges of multi-task learning by incorporating a Mixture-of-Experts (MoE) architecture. This innovative approach explicitly models the relationship between individual tasks and employs a gating network to optimize the outcomes. In the context of our task, we aim to use the MoE module to simultaneously capture both query-independent semantic information and query-shared information. These captured representations are subsequently fused together to enhance the overall robustness.

For each agent role, we construct an independent ``expert'' network to capture agent-specific representations, along with an agent-shared "expert" network that captures representations shared across different agents. To achieve this, we employ an adapter structure for each expert in our approach, drawing inspiration from the work~\cite{houlsby2019parameter}. Adapter tuning, a widely adopted strategy in natural language processing, involves updating only a minimal number of parameters while achieving substantial improvements in generalization across various downstream tasks~\cite{wang2020k, li2023hamur}. The adapter modules adhere to a bottleneck design, encompassing a down-projection layer, a non-linear layer, an up-projection layer, and a skip connection. This module seamlessly integrates as a plug-in directly within the transformer architecture.

In our model, for each agent, we construct an agent-independent adapter $\mathcal{A}^d_k$ cell, where $k \in [0,...,4]$. Additionally, an agent-shared adapter cell is formulated as $\mathcal{A}^s$ to capture shared information across distinct queries. Mathematically, we can express this as follows

\begin{equation}
\begin{aligned}
 \boldsymbol{v}^k_i &=  \mathcal{A}^d_k(\boldsymbol{e}^k_i)\\
 \boldsymbol{w}^k_i &= \mathcal{A}^s(\boldsymbol{e}^k_i)
 \end{aligned}
\end{equation}
where $\boldsymbol{v}^k_i$ represents the agent-independent representation for the $k$-th adapter and $i$-th query, and $\boldsymbol{w}^k_i$ represents the agent-shared representation, with all different agents share the same adapter cell $\mathcal{A}^s$. When $k$ is set 0, t specifically refers to the processing of the original query. The integration of these representations is efficiently executed through a gate network.

The gate network is designed to balance the proportion of individual and shared components in the model. The aforementioned operations decompose the representation $\boldsymbol{e}_i$ into independent embeddings $\boldsymbol{v}_i$ and shared embeddings $\boldsymbol{w}_i$, with the shared embedding capturing common semantic information. Reintegrating these components via the gating network effectively enhances the model's overall robustness. The gating network is designed as an MLP module. The fusion process is specified as follows
\begin{equation}
 \boldsymbol{g}^k_i =  \text{softmax}(W^k\boldsymbol{e}^k_i)
\end{equation}
where $\boldsymbol{W}^k \in \mathbb{R}^{2\times d}$ is a trainable matrix of gating network, $d$ denotes the feature dimension, then we fuse the $v^k_i$ and $w^k_i$ through gate net, expressed as
 \begin{equation}
\boldsymbol{h}^k_i =  \boldsymbol{g}^k_i[\boldsymbol{v}^k_i,\boldsymbol{w}^k_i]
\end{equation}
where $[\cdot]$ denotes the concat operation, $h^k_i$ is the aggregation result. Then, we will map the fusing results $h^k_i$ through a classier, denoted as
\begin{equation}
\boldsymbol{\hat{y}}^k_i = \mathcal{C}(\boldsymbol{h}^k_i)
\end{equation}
where $\mathcal{C}$ is a classifier, as we will normalize and classify the result into different classification levels。

\subsubsection{Loss Design}

In this part, we propose a novel loss function that concurrently optimizes accuracy and enhances model robustness by incorporating specific constraints into the loss function itself. Our design comprises two integral components: the accuracy loss and the robustness loss.

To address the challenge of accuracy loss, we employ the pointwise cross-entropy loss function, calculated on an individual query-document pair basis, with each pair treated independently. Within our experimental setup, the industrial dataset comprises labels classified into five relevance levels, contrasting with the three-level relevance classification of the public dataset we utilize. To harmonize these differing label structures, we initially transform the original value labels into a format compatible with a one-hot encoding scheme, denoted as $\hat{y}$. This step is followed by the application of cross-entropy calculations to juxtapose the predicted labels against the ground truth labels. The detailed procedure for this calculation is outlined below.

In addressing the issue of accuracy loss, we opt for the pointwise cross-entropy loss. This particular loss function is computed on a query-document pair basis, treating each pair independently. In our experimental framework, our industrial dataset contains labels originally categorized into five levels of relevance, whereas the public dataset we employ utilizes a three-level relevance classification. Mathematically, the calculation process is depicted as follows

Consequently, in the initial phase, we encode the original value labels to correspond to a one-hot encoding scheme, denoted as $\boldsymbol{y}^k_i$. Subsequently, we employ the cross-entropy calculation to compare the predicted labels with the ground truth labels. The calculation process can be elaborated as follows
\begin{equation}
 \mathcal{L}_{\text{acc}} = -\frac{1}{N}\frac{1}{K} \sum_{i}^{N}\sum_{k}^{K} (\boldsymbol{y}^k_i \cdot \log \boldsymbol{\hat{y}}^k_i + (1 - \boldsymbol{y}^k_i) \cdot \log(1 - \boldsymbol{\hat{y}}^k_i)) 
\end{equation}
here, $\boldsymbol{y}^k_i$ signifies the golden label of datasets, while $\hat{y}^k_i$ denotes the predicted label for the $i$-th query and the $k$-th agent. Here, $N$ and $K$ represent the total number of training samples and the aggregate number of agents, respectively. The loss function is specifically designed to calculate the logarithmic discrepancies between the predicted labels and actual relevance labels. This metric acts as a gauge for the model's accuracy within the scope of point-wise ranking.

In addition to addressing accuracy loss, it is imperative to consider the robustness of ranking performance, which constitutes a primary focus of this paper. To enhance robustness, our design strategy involves minimizing the divergence among different agents. While previous research like~\cite{zhuang2022characterbert,tasawong2023typo} have incorporated KL divergence in the loss function to reduce the distribution disparity between two predictions, this approach is not without limitations. Specifically, KL divergence is an asymmetric metric, potentially leading to an unbalanced problem. Given that our scenario involves multiple agent groups, where each agent is expected to maintain an equal status, we have selected the Jensen-Shannon (JS) divergence as a more equitable metric, denoted as
\begin{equation}
\mathcal{L}_{JS}(\boldsymbol{y}^a,\boldsymbol{y}^b) = \frac{1}{2}\mathcal{L}_{KL}(\boldsymbol{y}^a,\boldsymbol{y}^b) + \frac{1}{2}\mathcal{L}_{KL}(\boldsymbol{y}^b,\boldsymbol{y}^a) 
\end{equation}
then we use JS divergence to minimize the distribution gap among all the agent pair distributions, formulated as follows
\begin{equation}
\mathcal{L}_{\text{rbt}} = -\frac{1}{N} \sum_{m}^{K}\sum_{n}^{K} \mathcal{L}_{JS}( \boldsymbol{\hat{y}}^m_i,\boldsymbol{\hat{y}}^n_i), m<n
\end{equation}
minimizing $\mathcal{L}_{\text{rbt}}$ will reduce the discrepancy between each agent pair distribution. This way, we implicitly align representations of each agent and satisfy the robustness property.

Then we combine $\mathcal{L}_{\text{rbt}}$ and $\mathcal{L}_{\text{acc}}$ to get the total loss
\begin{equation}
\label{equ:Loss}
\mathcal{L}_{\text{total}}  = \mathcal{L}_{\text{acc}} + \alpha \mathcal{L}_{\text{rbt}} 
\end{equation}
The parameter $\alpha$ represents a tunable hyperparameter, instrumental in balancing between accuracy and robustness loss. Through this synergistic combination of losses, we successfully attain all the targeted properties of our model.

%% file: 4Experiment.tex
\section{Experiment}
We conduct thorough experiments to assess our proposed framework. The section will begin by describing the dataset, outlining the baseline models, and specifying the metrics employed for evaluation. Following this, we will delve into the evaluation of queries and the examination of model performance. Additionally, we will present results for ablation experiments and the impact of hyperparameters.

\subsection{Datasets Description}
We conduct experiments on two datasets: one public dataset Robust04 and one industrial dataset from Baidu.
\begin{itemize}[leftmargin=*]
  \item \textbf{Robust04}\footnote{https://trec.nist.gov/data/robust/04.guidelines.html}:
  The Robust04 dataset was specifically developed for the TREC 2004 Robust Track, a component of the Text Retrieval Conference (TREC) series. This dataset comprises 249 queries, each associated with relevance judgments on a collection of 528K documents.
  \item \textbf{Industrial Dataset}: 
  The industrial dataset we employed is a Chinese language dataset, which was sourced from user video search data collected from Baidu, the largest search engine platform in China. In this dataset, the queries and documents were extracted from real-world search logs, and the corpus corresponds to the documents generated through OCR recognition of the associated video URLs. The dataset comprises 16K queries and 430K documents. 
\end{itemize}

\subsection{Baseline}
During our experiment, we use the following baselines:

\begin{itemize}[leftmargin=*]
  \item \textbf{BM25}~\cite{robertson2009probabilistic}:
  BM25 is a straightforward yet powerful bag-of-words retrieval model that calculates relevance scores based on query term frequency, inverse document frequency, and document length.
  \item \textbf{ANCE}~\cite{xiong2020approximate}:
  ANCE uses an approximate nearest neighbor index for hard negative sample selection, integrated with model fine-tuning for dynamic training updates, leading to faster convergence, improved retrieval performance, and efficient utilization of computational resources. 
\end{itemize}

\subsection{Evaluation Metric}
To comprehensively assess our results, we employed both effectiveness and robustness evaluation metrics.
\subsubsection{Effectiveness Metric}
For effectiveness, we select the NDCG@N and MAP to evaluate effectiveness.
\subsubsection{Robustness Metric}
Following the previous work~\cite{wu2022neural}, we select the Variance of Normalized Average Precision (VNAP) as a robustness evaluation metric. Furthermore, we introduce the Variance of Normalized Discounted Cumulative Gain (VNDCG) for a comprehensive joint assessment. We use these two metrics to evaluate robustness performance across different agents. Mathematically, these metrics can be elaborated as follows
\begin{equation*}
NAP(q) = \frac{AP(q)}{E(AP(q))}
\end{equation*}
\begin{equation}
VNAP = \mathbb{V}(NAP_1(q),...,NAP_K(q))  
\end{equation}
where AP denotes the average precision with respect to the query $q$, NAP denotes the normalized average precision, and VNAP denotes the variance of a list NAP across different $K$ agents, and for VNDCG, it is defined as follows

\begin{equation}
VNDCG@N = \mathbb{V}(NDCG_1@N,...,NDCG_K@N)
\end{equation}
where NDCG@N represents the normalized discounted cumulative gain at $N$, while VNDCG@N denotes the variance observed across a series of NDCG@N values from different $K$ agents.

\subsection{Implementation Details}

In our experiment, we utilized various pre-trained models to executive~\eqref{equ:PLM}, including BERT, ERNIE, and RoBERTa, which serve as our base models. Specifically, for BERT-CE\footnote{https://huggingface.co/bert-base-chinese}\footnote{https://huggingface.co/bert-base-uncased}, ERNIE-CE\footnote{https://huggingface.co/nghuyong/ernie-3.0-base-zh}\footnote{https://huggingface.co/nghuyong/ernie-2.0-base-en}, and RoBERTa-CE\footnote{https://huggingface.co/hfl/chinese-roberta-wwm-ext}\footnote{https://huggingface.co/deepset/roberta-base-squad2}, we used different checkpoints from Hugging Face for the two datasets according to their language types, then finetuned them as cross-encoders on training datasets. Our experimental approach, such as in BERT+ours, adheres to a two-stage paradigm. In the first stage, we load the BERT-CE PLM checkpoint, followed by freezing the pre-trained model in the second stage. we integrate the insert PLM into the robust ranking structure and execute fine-tuning tasks on the training dataset. All experiments were conducted on a single Tesla V100 32G GPU using the AdamW optimizer. For baseline comparisons, we employed the open-sourced BM25 implementation from BEIR~\cite{thakur2021beir}. Our implementation of ANCE utilizes RoBERTa as the base pre-trained language model. Regarding CharacterBert, we follow the configuration in~\cite{zhuang2022characterbert}, with a modification with a point-wise loss function with align to our setting for a fair comparison.

\subsection{Query Quality Experiment}
Evaluating rewritten query results is crucial, as they significantly impact overall performance. Current assessment challenges arise from the alteration of semantic information and the need to evaluate character traits, areas lacking robust evaluation methods. To address this, we employ large language models to assess the rewritten queries. Our evaluation focuses on two key metrics: the semantic fidelity of the rewritten queries and their alignment with the defined agent's roles. We use diverse, publicly LLMS for both Chinese and English query assessments. We design specific prompts for the evaluation, enabling them to rate each metric on a 0 to 5 scale, with 0 being the poorest and 5 being the optimal performance. 
\begin{figure}[ht]
  \centering
\includegraphics[width=0.485\linewidth]{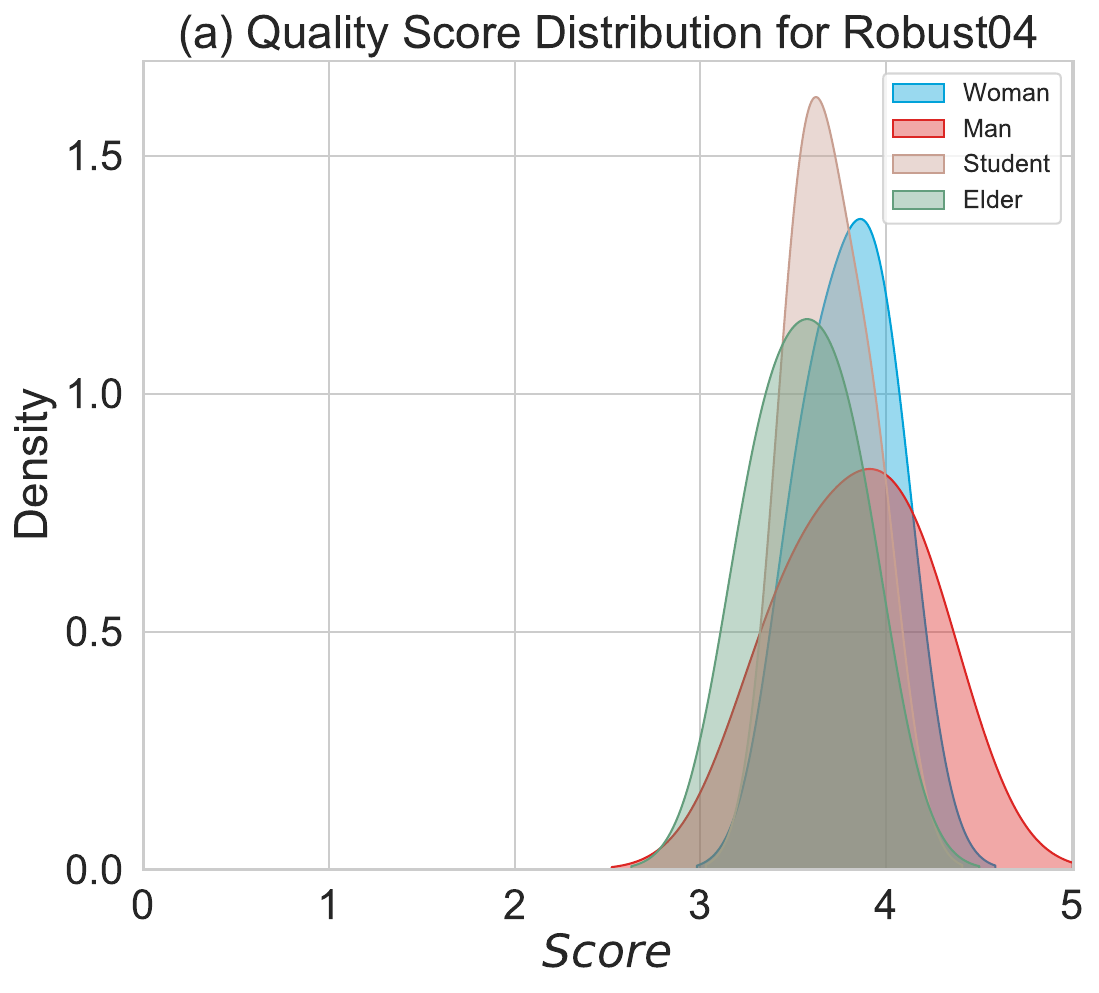}
 \includegraphics[width=0.485\linewidth]{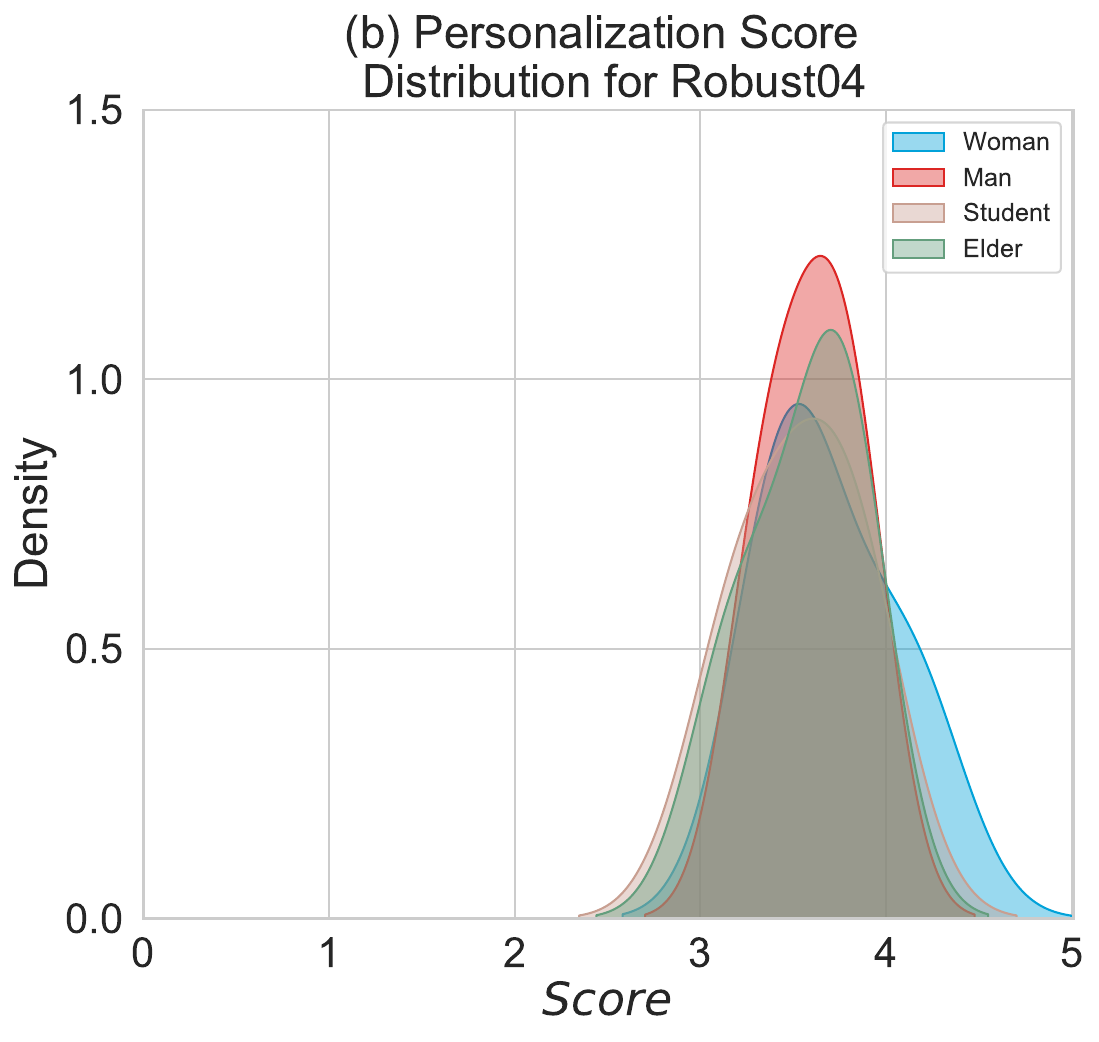}
 \includegraphics[width=0.485\linewidth]{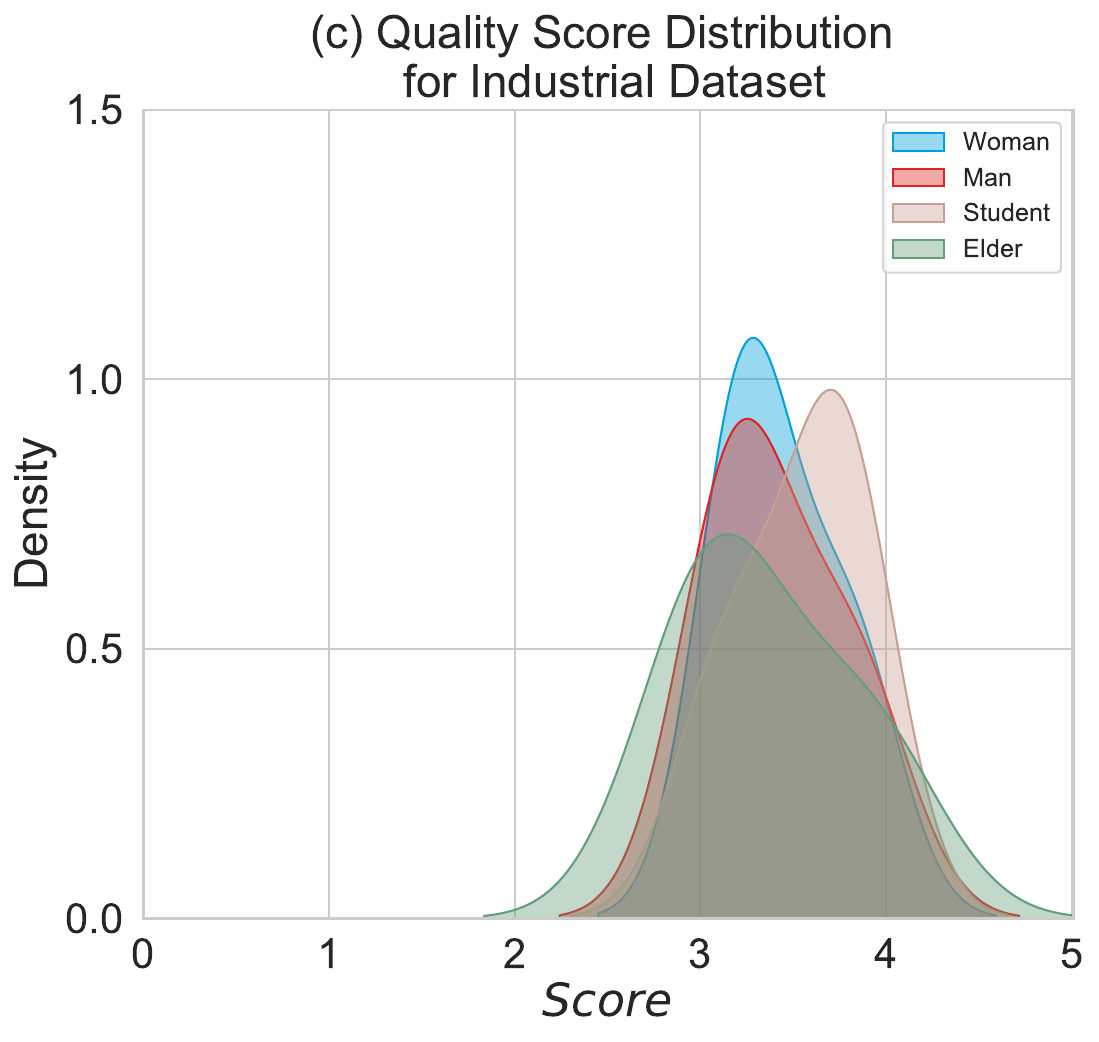}
\includegraphics[width=0.485\linewidth]{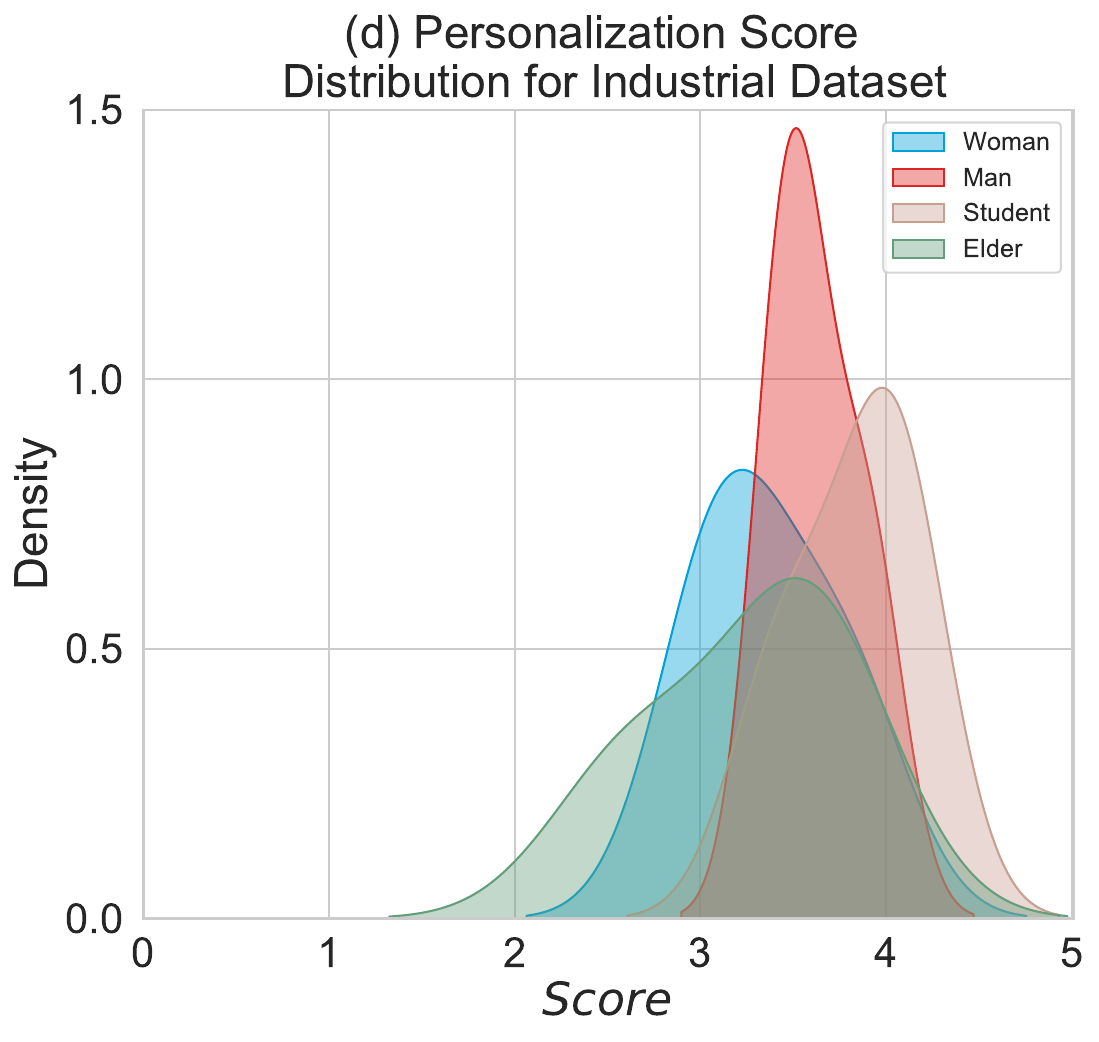}
 \caption{The Evaluation of Queries}
 \label{fig:evaluation_query}
 \vspace*{-3mm}
\end{figure}
As LLMs for Chinese, we select:
 \begin{itemize}[leftmargin=*]
\item \textbf{ChatGLM2-6B}\footnote{https://huggingface.co/THUDM/chatglm2-6b}: Tsinghua University's second-generation bilingual chat model in the ChatGLM series.
\item \textbf{Atom-7B}\footnote{https://huggingface.co/FlagAlpha/Atom-7B}: A joint development by Llama Chinese community, Atom-7B is based on Llama2-7B and further pre-trained with extensive Chinese data.
\item \textbf{ERNIE-Bot}\footnote{https://yiyan.baidu.com/}: Baidu's ERNIE Bot, a 2023 knowledge-enhanced generative AI, aims for accurate and fluent human-like responses, understanding human intentions effectively.
\end{itemize}
As LLMs for English, we select:
 \begin{itemize}[leftmargin=*]
  \item \textbf{vicuna-13B}\footnote{https://https//huggingface.co/lmsys/vicuna-13b-v1.3}:
  An open-source conversational model built by fine-tuning the Llama 13B model with user-shared conversations from ShareGPT.
  \item \textbf{Claude}\footnote{www.anthropic.com}: 
  A large, conversationally-focused language model developed by Anthropic.
  \item \textbf{Llama-2-13b}\footnote{https://huggingface.co/meta-llama/Llama-2-13b-chat-hf}:
  Meta and Microsoft's open-source Llama 2, for research and commercial use, features an optimized transformer architecture.
\end{itemize}

\begin{table*}[ht]
\begin{center}
 \caption{Performance Comparison Results.}
 \vspace*{-1mm}
 \label{tab:effectiveness}
 \setlength{\tabcolsep}{1mm}{
 \begin{tabular}{c|ccccc|ccccc}
  \hline
  \multirow{2}{*}{Models / \textbf{NDCG@10}} &\multicolumn{5}{c}{\textbf{Robust04}} &\multicolumn{5}{c}{\textbf{Industrial Dataset}}\\
   \cline{2-11}
   & \multicolumn{1}{c}{original} & \multicolumn{1}{c}{woman}& \multicolumn{1}{c}{man} & \multicolumn{1}{c}{student} &\multicolumn{1}{c|}{old}&\multicolumn{1}{c}{original} & \multicolumn{1}{c}{woman}& \multicolumn{1}{c}{man} & \multicolumn{1}{c}{student} &\multicolumn{1}{c}{old}\\
  \hline
  BM25            &0.4262 & 0.4062 & 0.3798 & 0.4259 & 0.3792 & 0.5380 & 0.5259 & 0.5264 & 0.5392 & 0.5183 \\
  ANCE            &0.4037 & 0.3905 & 0.3774 & 0.3802 & 0.3723 & 0.5640 & 0.5595 & 0.5588 & 0.5492 & 0.5587 \\
  CharacterBert   &0.4489 & 0.4192 & 0.4332 & 0.4274 & 0.4195 & 0.5860 & 0.5799 & 0.5859 & 0.5802 & 0.5733 \\
  BERT-CE         &0.4423 & 0.4129 & 0.4084 & 0.4082 & 0.4010 & 0.5821 & 0.5602 & 0.5637 & 0.5656 & 0.5679 \\
  ERNIE-CE       &0.4362 & 0.4021 & 0.3923 & 0.3993 & 0.4283 & 0.5888 & 0.5664 & 0.5687 & 0.5663 & 0.5699 \\
  RoBERTa-CE      &0.4562 & 0.4272 & 0.3997 & 0.4098 & 0.4194 & 0.5987 & 0.5740 & 0.5758 & 0.5809 & 0.5828 \\
  \hline
  \textbf{BERT+ours}   
  &0.4353 & 0.4102 & 0.4207 & 0.4178 & 0.4265 & 0.5924 & \textbf{0.5911*} & 0.5854 & 0.5821 & 0.5837 \\
  \textbf{ERNIE+ours}
  &0.4408 & 0.4305 & 0.4252 & 0.4365 & 0.4124 & 0.5896 & 0.5805 & 0.5829 & 0.5843 & 0.5870 \\ 
  \textbf{RoBERTa+ours} 
  & \textbf{0.4598*} & \textbf{0.4365*} & \textbf{0.4423*} & \textbf{0.4398*} & \textbf{0.4292*} & \textbf{0.5962*} & 0.5892 & \textbf{0.5879*} &\textbf{0.5843*} & \textbf{0.5877*}\\
 \hline
  \hline
  \multirow{2}{*}{Models / \textbf{MAP}} &\multicolumn{5}{c}{\textbf{Robust04}} &\multicolumn{5}{c}{\textbf{Industrial Dataset}}\\
   \cline{2-11}
   & \multicolumn{1}{c}{original} & \multicolumn{1}{c}{woman}& \multicolumn{1}{c}{man} & \multicolumn{1}{c}{student} &\multicolumn{1}{c|}{old}&\multicolumn{1}{c}{original} & \multicolumn{1}{c}{woman}& \multicolumn{1}{c}{man} & \multicolumn{1}{c}{student} &\multicolumn{1}{c}{old}\\
  \hline
BM25            & 0.2652 & 0.2528 & 0.2332 & 0.2643 & 0.2383 & 0.3584 & 0.3431 & 0.3571 & 0.3407 & 0.3391 \\
ANCE            & 0.2512 & 0.2453 & 0.2389 & 0.2592 & 0.2364 & 0.3753 & 0.3695 & 0.3698 & 0.3604 & 0.3674 \\
CharacterBert   & 0.2851 & 0.2623 & 0.2752 & 0.2589 & 0.2607 & 0.3950 & 0.3892 & 0.3932 & 0.3878 & 0.3821 \\
BERT-CE         & 0.2952 & 0.2694 & 0.2595 & 0.258 & 0.2572 & 0.3913 & 0.3797 & 0.3850 & 0.3874 & 0.3982 \\
ERNIE-CE       & 0.2864 & 0.2664 & 0.2659 & 0.2495 & 0.2722 & 0.4009 & 0.3876 & 0.3920 & 0.3968 & 0.3995 \\
RoBERTa-CE      & 0.2892 & 0.2467 & 0.2574 & 0.2503 & 0.2490 & 0.4081 & 0.3840 & 0.3894 & 0.3836 & 0.3855 \\
\hline
\textbf{BERT+ours}       
& 0.2887 & 0.2668 & 0.2792 & 0.2684 & \textbf{0.2809*} & 0.3947 & 0.3912 & 0.3896 & 0.3920 & 0.3974 \\
\textbf{ERNIE+ours}
& 0.2895 & \textbf{0.2858} & 0.2748 & 0.2793 & 0.2673 & 0.4072 & \textbf{0.4059*} & \textbf{0.4039*} & 0.3952 & 0.4018 \\
\textbf{RoBERTa+ours}
& \textbf{0.2961*} & 0.2797 & \textbf{0.2849*} & \textbf{0.2809*} & 0.2704 & \textbf{0.4083*} & 0.3921 & 0.3904 &\textbf{0.4021*}& \textbf{0.4037*} \\
 \hline
\end{tabular}}
\\``\textbf{{*}}'' indicates significance level test $p<0.05$.
\end{center}
\vspace*{-7mm}
\end{table*}
For a fair evaluation, we draw 100 queries for each role from each dataset, and assessed them by using three LLMs as mentioned earlier. Figure~\ref {fig:evaluation_query} presents the score distribution results. The results suggest high quality in both rewritten and persona queries, as evidenced by scores predominantly exceeding 3 out of 5. Notably, the quality of rewritten queries in English surpassed that of Chinese. We hypothesize this disparity stems from ChatGPT's proficiency in English tasks, paralleling our findings in role quality assessment. Additionally, the distribution does not indicate any role preference in ChatGPT, given the relatively uniform distribution across different roles. These outcomes robustly validate the effectiveness of employing LLMs for query rewriting across varied roles.

\begin{table*}[htbp]
\begin{center}
 \caption{Robustness Performance Comparison Results.}
 \vspace*{-1mm}
 \label{tab:robustness}
 \setlength{\tabcolsep}{1mm}{
 \begin{tabular}{c|cc|cc}%
  \hline
   \multirow{2}{*}{ \textbf{Metric}} &\multicolumn{2}{c}{\textbf{Robust04}}& \multicolumn{2}{c}{\textbf{Industrial Dataset}}\\
   & VNDCG@10~(e-5)$\downarrow$ & VNAP $\downarrow$  & VNDCG@10~(e-5)$\downarrow$ & VNAP $\downarrow$ \\
  \hline
  BM25            & 43.53 &  1.1823 & 6.287 &  1.0372 \\
  ANCE            & 12.43 &  0.9827 & 2.336 &  0.8764  \\
  CharacterBert   & \underline{12.01} &  \underline{0.9723} & \underline{2.209} & \underline{0.8394}  \\
  BERT-CE         & 20.69 &  1.0212 & 5.625 &  1.0360  \\
  ERNIE-CE        & 29.96 &  1.0519 & 7.210 &  1.0142   \\
  RoBERTa-CE      & 36.97 &  1.0745 & 7.635 &  1.0284  \\
  \hline
  \textbf{BERT+ours}            & \textbf{8.841*}(-26.3\%) &  \textbf{0.9082*}(-6.6\%)  & 1.650(-25.3\%) &  0.7176(-14.4\%)  \\
  \textbf{ERNIE+ours}           & 9.754(-16.9\%) &  0.9242(4.9\%)  & 1.518(-31.2\%) &  0.6350(-11.5\%)  \\
  \textbf{RoBERTa+ours}         & 10.29(-14.3\%) &  0.9473(2.6\%)  & \textbf{1.427*}(-35.4\%) &  \textbf{0.6165*}(-26.5\%)  \\
  \hline
\end{tabular}}
\\ The optimal results are highlighted in \textbf{bold}, the suboptimal results are \underline{underlined}. $\downarrow$: the lower the better. \\``\textbf{{*}}'' indicates significance level test $p<0.05$.
\end{center}
\vspace*{-7mm}
\end{table*}
\subsection{Performance Experiment}
In this section, we will display the result performance, including effectiveness and robustness, on two datasets. We compared our results against various baseline models, and we incorporated three transformer-based cross-encoders for aomparasion: BERT-CE, ERNIE-CE, and RoBERTa-CE. In our model, we adopted the same three transformer backbones, integrating our robust enhancing module atop them. We documented the results for all roles across both public and industrial datasets. Tables~\ref{tab:effectiveness} and ~\ref{tab:robustness} display the effectiveness and robustness findings respectively. 

\subsubsection{Effevtivess Analysis}
In this part, we will first discuss the performance impact of our model for the final result.
\begin{itemize}[leftmargin=*]
  \item When comparing our model with BM25 and ANCE, it is evident that transformer-based models, like ours, yield superior results due to their enhanced capability in capturing sentence relationships. Further comparison with CharacterBert, which is grounded in a self-teaching paradigm, reveals its proficient ranking ability. Our model excels in performance, particularly due to the systematic effect of our loss design and the MMoE-adapter network.
  \item Comparing rewritten queries with original query results, it is observed that all models, except CharacterBert, show inferior performance. This inferiority stems from their lack of robustness in handling out-of-distribution datasets. However, both CharacterBert and our model, designed with a robust loss function, successfully impose constraints across varied agent roles, ensuring consistent performance across different rewritten queries.
  \item When comparing our model with cross-encoders using the original datasets, we occasionally observe a slight decline in performance for our model. This can be reasonably attributed to potential semantic differences between queries. Consequently, our approach maintains a careful equilibrium between accuracy and robustness. To enhance robustness, a minor compromise in efficiency is sometimes necessary. However, in most cases, our model demonstrates improved accuracy. This improvement is facilitated by an additional stage of fine-tuning aimed at boosting model effectiveness. Furthermore, our uniquely designed loss function incorporates an accuracy component, ensuring continual improvement in accuracy.
  
\end{itemize}

\subsubsection{Robustness Analysis}
Robustness is our main concern in this paper, and the results are shown in Table~\ref{tab:robustness}. The analysis is listed as follows
\begin{itemize}[leftmargin=*]
  \item Our model achieves the highest performance, owing to its framework design that effectively captures semantic similarities across different groups and employs a loss function to unify these groups. BM25 exhibits lower robustness due to its nature of fragile attributes for statistical models, vulnerable to varied queries. BERT-CE, ERNIE-CE, and RoBERTa-CE also show poor robustness, lacking generalization capabilities across various queries. In contrast, ANCE models fare better, attributed to their neighborhood search process that inherently adds robustness. CharacterBert models, implementing a self-teaching process and leveraging KL-divergence, significantly narrow performance gaps. Our models utilize JS-divergence in loss design and employ a multi-expert adapter network, dynamically addressing commonalities in agent queries, thereby outperforming all other models.
  \item In comparisons with various cross-encoders on both datasets, our models consistently indicate an enhancement in overall robustness performance. This improvement can be attributed to our innovative plug-in module design, during which we employ two stages of fine-tuning to more effectively improve the robustness of existing models. However, due to the differing parameter scales and capabilities of various backbone models, performance variations are observed. Notably, BERT-based backbone models demonstrate the most substantial performance improvements in Robust04, while RoBERTa-based models excel in industrial datasets. This difference could be attributed to different data distribution.
\end{itemize}

\subsection{Ablation Experiment}
The ablation study results are detailed in Table~\ref{tab:ablation}. In this study, we evaluate five metrics on the industrial dataset: two effectiveness metrics (NDCG@10 and NDCG@20) and three robustness metrics (VNDCG@10, VNDCG@20, and VNAP). Our evaluation considers three distinct settings: first, the removal of the robust loss component, retaining only the accuracy loss (denoted as w/o-L); second, the exclusion of the robust MMoE module (denoted as w/o-N); and third, the elimination of both two components (denoted as w/o-N+L). Additionally, BERT is employed as the backbone model for this experiment.

Our findings indicate that our model consistently outperforms across all experiments. However, there are additional observations that warrant discussion. Firstly, an analysis of the results reveals that the use of MMoE introduces certain interferences in effectiveness, particularly when compared with the results from w/o-N+L and w/o-N configurations. This suggests that relying solely on MMoE modules may be counterproductive due to potential negative impacts on the final outcomes. Moreover, when examining robustness metrics, both the robust loss and MMoE modules appear to play a crucial role in enhancing robustness. However, the robust loss seems to have a more significant impact on robustness. We hypothesize that this is due to its direct influence on the model's performance, making it a pivotal factor in achieving superior results.

\begin{table}[htbp]
\begin{center}
 \caption{Ablation Experiment Results On Industrial Dataset.}
 \vspace*{-1mm}
 \setlength{\tabcolsep}{1mm}{
 \label{tab:ablation}
 \begin{tabular}{c|c|ccc}%
  \hline
   & \textbf{Ours} & \textbf{w/o-L} &\textbf{w/o-N} &\textbf{w/o-N+L}\\
  \hline
  NDCG@10 & \textbf{0.5881*} & 0.5725 & 0.5847 & 0.5739 \\
  NDCG@20 & \textbf{0.6239*} & 0.6102 & 0.6130 & 0.6109 \\
  VNDCG@10(e-5) &\textbf{1.650*} & 4.795 & 2.206 & 5.625 \\
  VNDCG@20(e-5) &\textbf{1.903*} & 5.143 & 2.594 & 7.372 \\
  VNAP & \textbf{0.7176*} & 0.9068 & 0.8346 & 1.0360 \\
  \hline
\end{tabular}}
\\``\textbf{{*}}'' indicates significance level test $p<0.05$.
\end{center}
\vspace*{-7mm}
\end{table}
\subsection{Hyper Parameter Experiment}

\begin{figure}[ht]
  \centering
 \includegraphics[width=0.485\linewidth]{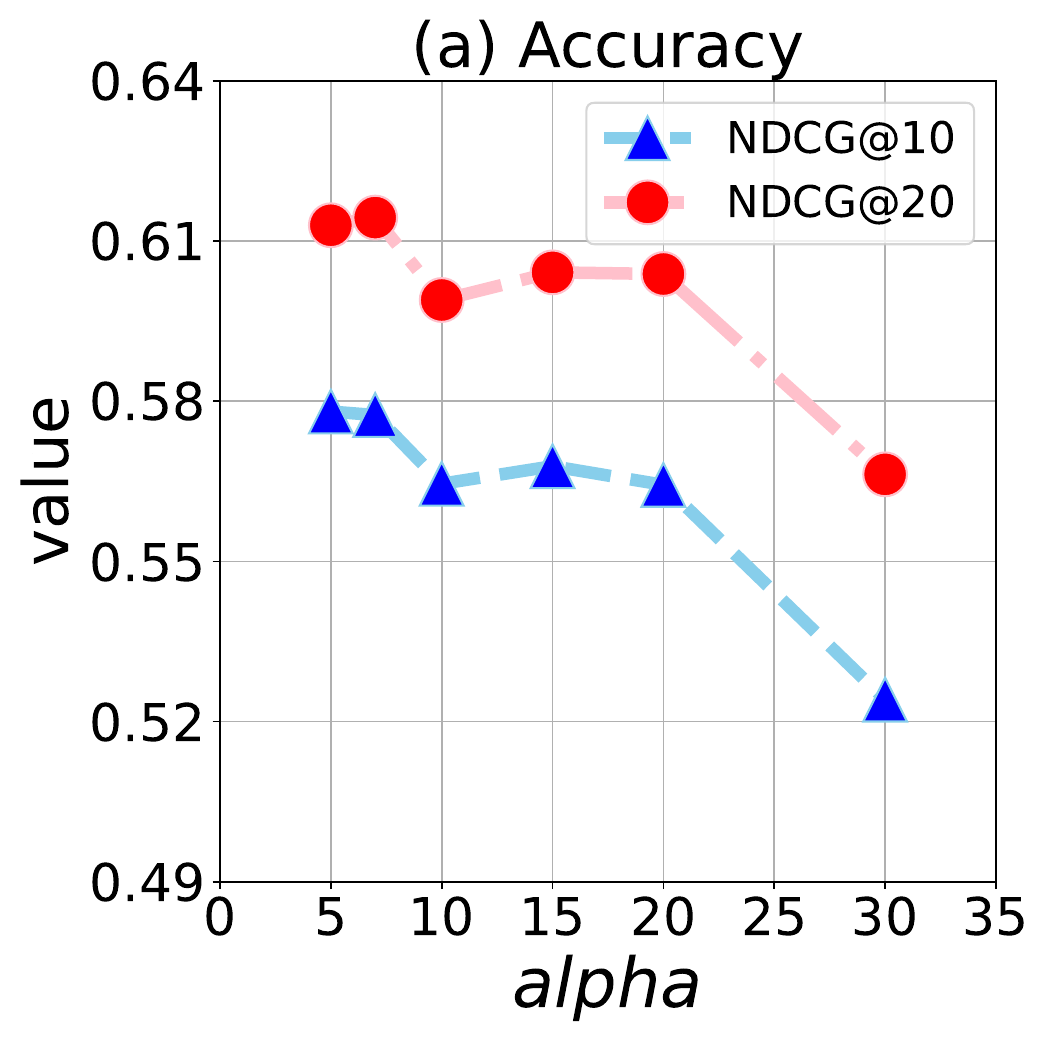}
 \includegraphics[width=0.485\linewidth]{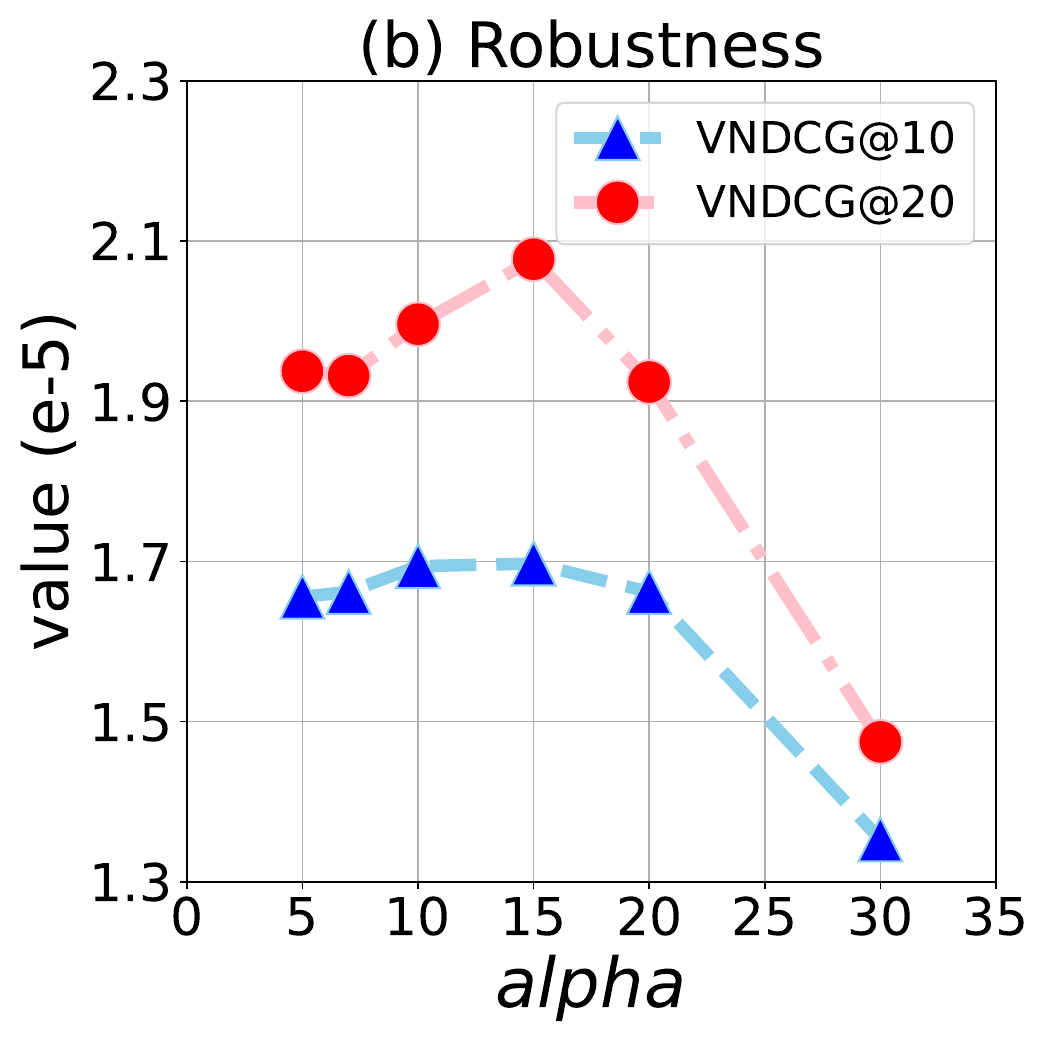}
 
 \caption{Hyper-parameters experiment}
 \label{fig:hyper}
 \vspace*{-3mm}
\end{figure}

In this part, we explore the impact of the hyper-parameter $\alpha$ as defined in Equation~\eqref{equ:Loss}. The parameter $\alpha$ is pivotal, determining the balance between accuracy and robust loss within the overall loss, and consequently, it has a direct influence on the model's final performance. Extending our previous discussion, we examine how changes in $\alpha$ impact the model's accuracy and robustness. An  $\alpha$ range from 5 to 30 was methodically chosen to study the resulting variations in accuracy and robustness metrics. These outcomes are depicted in Figure~\ref{fig:hyper}. As 
$\alpha$ is scaled up, the proportion of robust loss increases, leading to a decrease in the proportion of accuracy loss. Consequently, we observe a general uptrend in robustness while accuracy exhibits a downtrend. However, as $ \alpha$ becomes larger, we notice a downtrend in both accuracy and robustness. Thus, the reasonable selection of $\alpha$ is crucial for the model's optimal performance, which should consider both robustness and accuracy.

%% file: 5Relatedwork.tex
\section{Related Work}
During this section, an introduction related to our topic will be elaborated.
We will first introduce the recent research on query rewriting for large language models. Then, we will present recent studies on robust ranking.
\subsection{LLMs for Query Rewriting}

Traditional search engine systems encounter difficulties with ambiguous user queries, resulting in a vocabulary mismatch. The emergence of Large Language Models (LLMs) offers a substantial opportunity to revolutionize query rewriting capabilities. We categorized current LLMs-based query rewriting methods into four types: The first type is the prompting strategy, which deging prompt to instruct LLMs to rewrite queries. The approach like zero-shot prompting~\cite{feng2023knowledge,mackie2023generative,shen2023large, alaofi2023can}, few-shot learning~\cite{jagerman2023query, wang2023query2doc}, and CoT prompting, like~\cite{jagerman2023query}. 
The second type is the fine-tuning strategy, which fine-tune LLMs on on specific datasets. in the paper~\cite{ma2023query}, fine-tuned LLMs are employed to modify queries, and then read the associated documents, contributing to the refinement of query quality.
The third category is retrieval-augmented methods, involving rewriting queries using retrieval strategies. In \cite{ma2023query}, an external search engine supplements query information. Similarly, \cite{wang2023query2doc} describes creating fake documents and employing a web search engine to obtain external documents as ground-truth data for query rewriting. 
Another type is the knowledge distillation method, addressing the deployment limitations of LLMs due to reference speed constraints. In \cite{srinivasan2022quill}, the author first fine-tuned an LLM on the query-rewriting dataset, followed by two steps of distillation. Firstly, a professor model is distilled into a teacher model, and then, the teacher model is distilled into a lightweight BERT student model for query rewriting.
Our method falls into prompting strategies. However, it distinctively utilizes LLMs as varied role agents, diverging from traditional prompting approaches.

\subsection{Robust Ranking}

Pre-trained language models have demonstrated exceptional performance in retrieval and ranking tasks. However, while traditional research has predominantly focused on effectiveness, neural models are often susceptible to errors when confronted with altered input, prompting recent studies to prioritize robustness.
In~\cite{penha2022evaluating}, the author assesses the robustness of current retrieval pipelines by introducing various query variants, including specialization, misspelling, naturality, ordering, and paraphrasing. The findings reveal a significant decrease in effectiveness when queries undergo disturbances.
In \cite{campos2023noise}, a contrastive aligning method is proposed to enhance robustness. This method clusters positive samples, distinguishing them from negative samples to improve overall system robustness. Another approach, CharacterBert~\cite{zhuang2022characterbert}, is introduced to bolster robustness in the face of query typos. CharacterBert is used to encode both the original document and queries, employing a self-teaching paradigm with KL-divergence to enforce an identical distribution across queries and queries with typos.
In \cite{lupart2023study}, adversarial training is proposed as a method to increase robustness. The authors explore FGSM techniques to make the model more robust to variations in input.
Addressing misspelling queries, work~\cite{tasawong2023typo} focuses on query augmentation models to generate misspelling queries. It calculates the similarity score distribution between each query and a passage list, then introduces a dual self-teaching loss for alignment. These studies collectively contribute to the ongoing efforts to enhance the robustness of pre-trained language models in retrieval tasks.

Our architecture for enhancing robustness incorporates two distinct tasks: the MMoE module and the loss design module. In contrast to self-teaching methods, our approach leverages JS divergence to narrow the gap between different distributions.

%% file: 6Conclusion.tex
\section{Conclusion}
In this paper, we introduce an innovative pipeline designed to enhance the robustness of existing ranking models in search engines from demographic perspectives. Our pipeline comprises two main components: query rewriting leveraging multi-role LLM agents and a novel framework including a MMoE robustness model and a robust loss function. During the initial phase, we employ LLMs in various roles as agents to effectively rewrite queries. Subsequently, we introduce a framework with an MMoE robust model and a robust loss function. These two elements jointly improve the robustness of the ranking results, particularly when dealing with queries that have different semantic inputs. Extensive experimentations are conducted to validate the effectiveness of our proposed framework. Future research will explore the possibility of directly involving LLMs in the ranking process and enhancing interpretability to further improve robustness.

%% file: 0Paperid1633.bbl
\begin{thebibliography}{10}
\providecommand{\url}[1]{#1}
\csname url@samestyle\endcsname
\providecommand{\newblock}{\relax}
\providecommand{\bibinfo}[2]{#2}
\providecommand{\BIBentrySTDinterwordspacing}{\spaceskip=0pt\relax}
\providecommand{\BIBentryALTinterwordstretchfactor}{4}
\providecommand{\BIBentryALTinterwordspacing}{\spaceskip=\fontdimen2\font plus
\BIBentryALTinterwordstretchfactor\fontdimen3\font minus
  \fontdimen4\font\relax}
\providecommand{\BIBforeignlanguage}[2]{{%
\expandafter\ifx\csname l@#1\endcsname\relax
\typeout{** WARNING: IEEEtran.bst: No hyphenation pattern has been}%
\typeout{** loaded for the language `#1'. Using the pattern for}%
\typeout{** the default language instead.}%
\else
\language=\csname l@#1\endcsname
\fi
#2}}
\providecommand{\BIBdecl}{\relax}
\BIBdecl

\bibitem{ntoulas2004s}
A.~Ntoulas, J.~Cho, and C.~Olston, ``What's new on the web? the evolution of
  the web from a search engine perspective,'' in \emph{Proceedings of the 13th
  international conference on World Wide Web}, 2004, pp. 1--12.

\bibitem{salton1975vector}
G.~Salton, A.~Wong, and C.-S. Yang, ``A vector space model for automatic
  indexing,'' \emph{Communications of the ACM}, vol.~18, no.~11, pp. 613--620,
  1975.

\bibitem{robertson1976relevance}
S.~E. Robertson and K.~S. Jones, ``Relevance weighting of search terms,''
  \emph{Journal of the American Society for Information science}, vol.~27,
  no.~3, pp. 129--146, 1976.

\bibitem{dai2019deeper}
Z.~Dai and J.~Callan, ``Deeper text understanding for ir with contextual neural
  language modeling,'' in \emph{Proceedings of the 42nd international ACM SIGIR
  conference on research and development in information retrieval}, 2019, pp.
  985--988.

\bibitem{dai2018convolutional}
Z.~Dai, C.~Xiong, J.~Callan, and Z.~Liu, ``Convolutional neural networks for
  soft-matching n-grams in ad-hoc search,'' in \emph{Proceedings of the
  eleventh ACM international conference on web search and data mining}, 2018,
  pp. 126--134.

\bibitem{guo2016deep}
J.~Guo, Y.~Fan, Q.~Ai, and W.~B. Croft, ``A deep relevance matching model for
  ad-hoc retrieval,'' in \emph{Proceedings of the 25th ACM international on
  conference on information and knowledge management}, 2016, pp. 55--64.

\bibitem{gu2020speaker}
J.-C. Gu, T.~Li, Q.~Liu, Z.-H. Ling, Z.~Su, S.~Wei, and X.~Zhu, ``Speaker-aware
  bert for multi-turn response selection in retrieval-based chatbots,'' in
  \emph{Proceedings of the 29th ACM International Conference on Information \&
  Knowledge Management}, 2020, pp. 2041--2044.

\bibitem{ma2021b}
X.~Ma, J.~Guo, R.~Zhang, Y.~Fan, Y.~Li, and X.~Cheng, ``B-prop: bootstrapped
  pre-training with representative words prediction for ad-hoc retrieval,'' in
  \emph{Proceedings of the 44th International ACM SIGIR Conference on Research
  and Development in Information Retrieval}, 2021, pp. 1513--1522.

\bibitem{dash2003consistency}
M.~Dash and H.~Liu, ``Consistency-based search in feature selection,''
  \emph{Artificial intelligence}, vol. 151, no. 1-2, pp. 155--176, 2003.

\bibitem{penha2022evaluating}
G.~Penha, A.~C{\^a}mara, and C.~Hauff, ``Evaluating the robustness of retrieval
  pipelines with query variation generators,'' in \emph{European conference on
  information retrieval}.\hskip 1em plus 0.5em minus 0.4em\relax Springer,
  2022, pp. 397--412.

\bibitem{campos2023noise}
D.~Campos, C.~Zhai, and A.~Magnani, ``Noise-robust dense retrieval via
  contrastive alignment post training,'' \emph{arXiv e-prints}, pp.
  arXiv--2304, 2023.

\bibitem{zhuang2022characterbert}
S.~Zhuang and G.~Zuccon, ``Characterbert and self-teaching for improving the
  robustness of dense retrievers on queries with typos,'' in \emph{Proceedings
  of the 45th International ACM SIGIR Conference on Research and Development in
  Information Retrieval}, 2022, pp. 1444--1454.

\bibitem{statistic2021cnnic}
CNNIC, ``Statistical report on internet development in china,''
  \url{https://www.cnnic.net.cn/n4/2022/0401/c88-1125.html}.

\bibitem{jia2023mill}
P.~Jia, Y.~Liu, X.~Zhao, X.~Li, C.~Hao, S.~Wang, and D.~Yin, ``Mill: Mutual
  verification with large language models for zero-shot query expansion,''
  \emph{arXiv preprint arXiv:2310.19056}, 2023.

\bibitem{hao2022cgf}
J.~Hao, Y.~Liu, X.~Fan, S.~Gupta, S.~Soltan, R.~Chada, P.~Natarajan, C.~Guo,
  and G.~T{\"u}r, ``Cgf: Constrained generation framework for query rewriting
  in conversational ai,'' in \emph{Proceedings of the 2022 Conference on
  Empirical Methods in Natural Language Processing: Industry Track}, 2022, pp.
  475--483.

\bibitem{gao2022precise}
L.~Gao, X.~Ma, J.~Lin, and J.~Callan, ``Precise zero-shot dense retrieval
  without relevance labels,'' \emph{arXiv preprint arXiv:2212.10496}, 2022.

\bibitem{wang2023query2doc}
L.~Wang, N.~Yang, and F.~Wei, ``Query2doc: Query expansion with large language
  models,'' \emph{arXiv preprint arXiv:2303.07678}, 2023.

\bibitem{shen2023large}
T.~Shen, G.~Long, X.~Geng, C.~Tao, T.~Zhou, and D.~Jiang, ``Large language
  models are strong zero-shot retriever,'' \emph{arXiv preprint
  arXiv:2304.14233}, 2023.

\bibitem{alaofi2023can}
M.~Alaofi, L.~Gallagher, M.~Sanderson, F.~Scholer, and P.~Thomas, ``Can
  generative llms create query variants for test collections? an exploratory
  study,'' in \emph{Proceedings of the 46th International ACM SIGIR Conference
  on Research and Development in Information Retrieval}, 2023, pp. 1869--1873.

\bibitem{jagerman2023query}
R.~Jagerman, H.~Zhuang, Z.~Qin, X.~Wang, and M.~Bendersky, ``Query expansion by
  prompting large language models,'' \emph{arXiv preprint arXiv:2305.03653},
  2023.

\bibitem{yu2022generate}
W.~Yu, D.~Iter, S.~Wang, Y.~Xu, M.~Ju, S.~Sanyal, C.~Zhu, M.~Zeng, and
  M.~Jiang, ``Generate rather than retrieve: Large language models are strong
  context generators,'' \emph{arXiv preprint arXiv:2209.10063}, 2022.

\bibitem{izacard2020leveraging}
G.~Izacard and E.~Grave, ``Leveraging passage retrieval with generative models
  for open domain question answering,'' \emph{arXiv preprint arXiv:2007.01282},
  2020.

\bibitem{ma2023query}
X.~Ma, Y.~Gong, P.~He, H.~Zhao, and N.~Duan, ``Query rewriting for
  retrieval-augmented large language models,'' \emph{arXiv preprint
  arXiv:2305.14283}, 2023.

\bibitem{anand2023query}
A.~Anand, A.~Anand, V.~Setty \emph{et~al.}, ``Query understanding in the age of
  large language models,'' \emph{arXiv preprint arXiv:2306.16004}, 2023.

\bibitem{reimers2019sentence}
N.~Reimers and I.~Gurevych, ``Sentence-bert: Sentence embeddings using siamese
  bert-networks,'' \emph{arXiv preprint arXiv:1908.10084}, 2019.

\bibitem{devlin2018bert}
J.~Devlin, M.-W. Chang, K.~Lee, and K.~Toutanova, ``Bert: Pre-training of deep
  bidirectional transformers for language understanding,'' \emph{arXiv preprint
  arXiv:1810.04805}, 2018.

\bibitem{liu2019roberta}
Y.~Liu, M.~Ott, N.~Goyal, J.~Du, M.~Joshi, D.~Chen, O.~Levy, M.~Lewis,
  L.~Zettlemoyer, and V.~Stoyanov, ``Roberta: A robustly optimized bert
  pretraining approach,'' \emph{arXiv preprint arXiv:1907.11692}, 2019.

\bibitem{sun2020ernie}
Y.~Sun, S.~Wang, Y.~Li, S.~Feng, H.~Tian, H.~Wu, and H.~Wang, ``Ernie 2.0: A
  continual pre-training framework for language understanding,'' in
  \emph{Proceedings of the AAAI conference on artificial intelligence},
  vol.~34, no.~05, 2020, pp. 8968--8975.

\bibitem{ma2018modeling}
J.~Ma, Z.~Zhao, X.~Yi, J.~Chen, L.~Hong, and E.~H. Chi, ``Modeling task
  relationships in multi-task learning with multi-gate mixture-of-experts,'' in
  \emph{Proceedings of the 24th ACM SIGKDD international conference on
  knowledge discovery \& data mining}, 2018, pp. 1930--1939.

\bibitem{houlsby2019parameter}
N.~Houlsby, A.~Giurgiu, S.~Jastrzebski, B.~Morrone, Q.~De~Laroussilhe,
  A.~Gesmundo, M.~Attariyan, and S.~Gelly, ``Parameter-efficient transfer
  learning for nlp,'' in \emph{International Conference on Machine
  Learning}.\hskip 1em plus 0.5em minus 0.4em\relax PMLR, 2019, pp. 2790--2799.

\bibitem{wang2020k}
R.~Wang, D.~Tang, N.~Duan, Z.~Wei, X.~Huang, G.~Cao, D.~Jiang, M.~Zhou
  \emph{et~al.}, ``K-adapter: Infusing knowledge into pre-trained models with
  adapters,'' \emph{arXiv preprint arXiv:2002.01808}, 2020.

\bibitem{li2023hamur}
X.~Li, F.~Yan, X.~Zhao, Y.~Wang, B.~Chen, H.~Guo, and R.~Tang, ``Hamur: Hyper
  adapter for multi-domain recommendation,'' in \emph{Proceedings of the 32nd
  ACM International Conference on Information and Knowledge Management}, 2023,
  pp. 1268--1277.

\bibitem{tasawong2023typo}
P.~Tasawong, W.~Ponwitayarat, P.~Limkonchotiwat, C.~Udomcharoenchaikit,
  E.~Chuangsuwanich, and S.~Nutanong, ``Typo-robust representation learning for
  dense retrieval,'' \emph{arXiv preprint arXiv:2306.10348}, 2023.

\bibitem{robertson2009probabilistic}
S.~Robertson, H.~Zaragoza \emph{et~al.}, ``The probabilistic relevance
  framework: Bm25 and beyond,'' \emph{Foundations and Trends{\textregistered}
  in Information Retrieval}, vol.~3, no.~4, pp. 333--389, 2009.

\bibitem{xiong2020approximate}
L.~Xiong, C.~Xiong, Y.~Li, K.-F. Tang, J.~Liu, P.~Bennett, J.~Ahmed, and
  A.~Overwijk, ``Approximate nearest neighbor negative contrastive learning for
  dense text retrieval,'' \emph{arXiv preprint arXiv:2007.00808}, 2020.

\bibitem{wu2022neural}
C.~Wu, R.~Zhang, J.~Guo, Y.~Fan, and X.~Cheng, ``Are neural ranking models
  robust?'' \emph{ACM Transactions on Information Systems}, vol.~41, no.~2, pp.
  1--36, 2022.

\bibitem{thakur2021beir}
N.~Thakur, N.~Reimers, A.~R{\"u}ckl{\'e}, A.~Srivastava, and I.~Gurevych,
  ``Beir: A heterogenous benchmark for zero-shot evaluation of information
  retrieval models,'' \emph{arXiv preprint arXiv:2104.08663}, 2021.

\bibitem{feng2023knowledge}
J.~Feng, C.~Tao, X.~Geng, T.~Shen, C.~Xu, G.~Long, D.~Zhao, and D.~Jiang,
  ``Knowledge refinement via interaction between search engines and large
  language models,'' \emph{arXiv preprint arXiv:2305.07402}, 2023.

\bibitem{mackie2023generative}
I.~Mackie, S.~Chatterjee, and J.~Dalton, ``Generative and pseudo-relevant
  feedback for sparse, dense and learned sparse retrieval,'' \emph{arXiv
  preprint arXiv:2305.07477}, 2023.

\bibitem{srinivasan2022quill}
K.~Srinivasan, K.~Raman, A.~Samanta, L.~Liao, L.~Bertelli, and M.~Bendersky,
  ``Quill: Query intent with large language models using retrieval augmentation
  and multi-stage distillation,'' \emph{arXiv preprint arXiv:2210.15718}, 2022.

\bibitem{lupart2023study}
S.~Lupart and S.~Clinchant, ``A study on fgsm adversarial training for neural
  retrieval,'' in \emph{European Conference on Information Retrieval}.\hskip
  1em plus 0.5em minus 0.4em\relax Springer, 2023, pp. 484--492.

\end{thebibliography}
